\documentstyle[12pt,aaspp4]{article}
\lefthead{Bureau et al.}
\righthead{The Dark Halo of NGC~2915}
\slugcomment{Accepted for Publication in The Astronomical Journal}
\begin{document}
%
%
\newcommand{\coltwo}[2]{\mbox{$\displaystyle\mbox{#1}\atop{\displaystyle\mbox{#2}}$}}
%
%
\title{The Shape and Figure Rotation of NGC~2915's Dark Halo}
\author{M.\ Bureau\altaffilmark{1}, K.\ C.\ Freeman, and D.\ W.\ Pfitzner}
\affil{Mount Stromlo and Siding Spring Observatories, Institute of Advanced
Studies, The Australian National University, Private Bag, Weston Creek P.O.,
ACT~2611, Australia}
\altaffiltext{1}{Now at Sterrewacht Leiden, Postbus 9513, 2300 RA Leiden, The
Netherlands}
\and
\author{G.\ R.\ Meurer}
\affil{Department of Physics and Astronomy, Johns Hopkins University,
Baltimore, MD 21218, U.S.A.}
\begin{abstract}
  NGC~2915 is a blue compact dwarf galaxy with a very extended \ion{H}{1}
  disk. This disk shows a short central bar and extended spiral arms, both
  reaching far beyond the optical component. We use Tremaine \& Weinberg's
  \markcite{tw84}(1984) method to measure the pattern speed of the bar and
  spiral arms from \ion{H}{1} radio synthesis data. Our measurements yield a
  pattern speed of $\Omega_p=0.21\pm0.06$~km~s$^{-1}$~arcsec$^{-1}$
  ($8.0\pm2.4$~km~s$^{-1}$~kpc$^{-1}$ for $D=5.3$~Mpc), in disagreement with
  the general view that corotation in barred disks lies just outside the end
  of the bar, but consistent with recent models of barred galaxies with dense
  dark matter halos. Our adopted bar semi-length $r_b\approx180\arcsec$ puts
  corotation at more than 1.7~$r_b$. The existence of the pattern is also
  problematic. Because NGC~2915 is isolated, gravitational interactions cannot
  account for the structure observed in the \ion{H}{1} disk. We also
  demonstrate that the low surface density observed in the disk and the
  location of the pseudo-rings make it unlikely that swing amplification
  (Toomre \markcite{t81}1981) or bar-driven spiral arms could explain the bar
  and spiral pattern.
  
  Based on the similarity of the dark matter and \ion{H}{1} surface density
  profiles, we discuss the possibility of dark matter distributed in a disk
  and following closely the \ion{H}{1} distribution. This disk then becomes
  gravitationally unstable and can naturally form a bar and spiral pattern.
  However, this explanation is difficult to reconcile with some properties of
  NGC~2915. Finally, we consider the effect of a massive and extended triaxial
  dark matter halo with a rotating figure. The existence of such halos is
  supported by CDM simulations showing strongly triaxial dark halos with slow
  figure rotation. The observed structure of the \ion{H}{1} disk can then
  arise through forcing by the rotating triaxial figure. We associate the
  measured pattern speed in NGC~2915 with the figure rotation of its dark
  halo.
\end{abstract}
\keywords{cosmology: dark matter~--- galaxies: evolution~--- galaxies:
formation~--- galaxies: individual: NGC~2915~--- galaxies: kinematics and
dynamics~--- galaxies: structure}
\section{Introduction\label{sec:introduction}}
\nopagebreak
About 30\% of spiral galaxies are strongly barred. This fraction is roughly
doubled if weaker bar-like asymmetries are included, and it reaches almost
unity if $K$-band images and quantitative methods are used for classification
(see, e.g., Seiger \& James \markcite{sj98}1998). Studying barred spiral
galaxies is therefore an important part of galactic structure and dynamics,
and we refer the reader to the review by Sellwood \& Wilkinson
\markcite{sw93}(1993).

The form of the potential and the bar pattern speed ($\Omega_p$) are the most
important quantities determining the structure and dynamics of a barred disk.
$N$-body simulations of bars (e.g.\ Sellwood \markcite{se81}1981; Combes \&
Sanders \markcite{cs81}1981) and analytical calculations (e.g.\ Contopoulos
\markcite{c80}1980; Contopoulos \& Grosb\o l \markcite{cg89}1989) have shown
that for bars to be self-consistent, their pattern speed should be such that
corotation occurs outside the end of the bar. However, measuring the pattern
speed of bars is not an easy task and most methods are rather indirect (see,
e.g., Schwarz \markcite{s81}1981, \markcite{s84}1984; Athanassoula
\markcite{a92b}1992b; Canzian \markcite{ca93}1993). The method proposed by
Tremaine \& Weinberg \markcite{tw84}(1984) is the only direct method, based on
observables alone. However, because of the assumptions it makes, the method is
rarely applicable to real galaxies and it has been applied successfully to
only two objects so far (NGC~936: Kent \markcite{k87}1987, Merrifield \&
Kuijken \markcite{mk95}1995; NGC~4596: Gerssen, Kuijken, \& Merrifield
\markcite{gkm99}1999). In this paper, we apply the Tremaine \& Weinberg (TW)
method to the galaxy NGC~2915, a blue compact dwarf (BCD) galaxy with a very
extended \ion{H}{1} disk (Meurer, Mackie, \& Carignan \markcite{mmc94}1994,
hereafter MMC94; Meurer et al.\ \markcite{mcbf96}1996, hereafter MCBF96). This
disk shows a late barred spiral galaxy morphology extending far beyond the
optical component.

While the formation of bars is now fairly well understood, the formation of
the spiral patterns seen in many galaxy disks remains an open issue. Various
mechanisms have been proposed to form and maintain spiral patterns. We will
consider the most prominent ones in this paper: gravitational interactions
(e.g.\ Noguchi \markcite{n87}1987), swing amplification of spiral density
waves (Toomre \markcite{t81}1981), and bar-driven spiral arms (e.g.\ Byrd et
al.\ \markcite{brsbc94}1994). We show that none of these is likely to explain
the bar and spiral arms structure observed in the \ion{H}{1} disk of NGC~2915.

This problem leads us to consider the background mass distribution in
NGC~2915. In order to reproduce their flat rotation curves, disks are commonly
believed to be immersed in massive dark halos (e.g.\ Carignan \& Freeman
\markcite{cf85}1985). Although the shape of such halos is often assumed to be
spherical, there is little observational information on this subject and other
shapes are clearly possible (see, e.g., Trimble \markcite{t87}1987 and Carr
\markcite{c94}1994 for reviews). In the case of NGC~2915 in particular, we
demonstrate that a simple rigid spherical halo is unable to explain the
properties of the \ion{H}{1} disk. We therefore consider other dark matter
distributions, including: 1) dark matter distributed in a thin disk following
the \ion{H}{1} surface density distribution, and 2) dark matter distributed in
an extended triaxial halo with a slowly rotating figure.

In \S~\ref{sec:pat_measure}, we discuss the TW method to measure the pattern
speed of bars and review its past applications. In \S~\ref{sec:ngc2915}, we
describe the properties of NGC~2915 and measure its pattern speed. The issues
of self-consistency and multiple pattern speeds are touched upon in
\S~\ref{sec:natpat}, and in \S~\ref{sec:oripat} we discuss the possible
formation mechanisms for the bar and spiral pattern. We consider two
particular models for the dark matter distribution of NGC~2915 in
\S~\ref{sec:dm}.
\section{Pattern Speed Measurement\label{sec:pat_measure}}
\nopagebreak
\subsection{Indirect Methods\label{sec:indirect}}
\nopagebreak
Many indirect methods have been devised to measure the bar pattern speed
$\Omega_p$ in barred spiral galaxies, and we discuss two of them here. The
first one relies on the theory of resonance ring formation (Schwarz
\markcite{s81}1981, \markcite{s84}1984; see also Byrd, Ousley, \& Dalla Piazza
\markcite{bod98}1998). It is possible to identify morphological features in
the disk of a barred galaxy, such as rings, with the locations of known
resonances. Using the circular velocity curve and derived epicyclic frequency,
the bar pattern speed can then be inferred (see, e.g., Ryder et al.\ 
\markcite{rbtssw96}1996 for recent applications to NGC~1433 and NGC~6300). We
apply this method to NGC~2915 in \S~\ref{sec:oripat}. Second, it is possible
to construct hydrodynamical models of the gas flow in a galaxy (often to
reproduce \ion{H}{1} observations because of their full kinematical coverage).
Hydrodynamical models depend on several parameters of the potential, including
$\Omega_p$, so the best fitting model gives an estimate of the bar pattern
(see, e.g., Lindblad \& Kristen \markcite{lk96}1996 and Lindblad, Lindblad, \&
Athanassoula \markcite{lla96}1996 for recent applications to NGC~1300 and
NGC~1365). Sellwood \& Wilkinson \markcite{sw93}(1993) review many such barred
galaxy studies.

Although these indirect methods yield useful information about pattern speeds,
it is clearly preferable to have a direct measurement, with no dependence on
theoretical arguments or comparison with dynamical models. We will thus pursue
these indirect methods only as a consistency check of more direct
measurements.
\subsection{Direct Methods\label{sec:direct}}
\nopagebreak
Tremaine \& Weinberg \markcite{tm84}(1984) showed that the pattern speed of a
barred disk galaxy can be derived from the observed kinematics of a tracer
population if (1) there exists a unique, well-defined disk pattern speed
$\Omega_p$, and (2) the luminosity density of the tracer satisfies the
continuity equation. With these conditions, the pattern speed is given by
\begin{equation}
\label{eq:tm}
\Omega_p\sin i=\frac{\int^{\infty}_{-\infty}h(y)dy\int^{\infty}_{-\infty}\Sigma(x,y)[\bar{v}_{los}(x,y)-v_{sys}]dx}
                    {\int^{\infty}_{-\infty}h(y)dy\int^{\infty}_{-\infty}\Sigma(x,y)xdx},
\end{equation}
where $i$ is the disk inclination, $\Sigma$ is the surface brightness
distribution of the tracer, $\bar{v}_{los}$ is the mean line-of-sight velocity
of the tracer, $v_{sys}$ is the systemic velocity of the galaxy, and $(x,y)$
are Cartesian coordinates on the plane of the sky, with the origin at the
center of the galaxy and the $x$-axis parallel to the line of nodes (i.e.\ the
apparent major axis for an intrinsically round disk). These quantities are all
observables. The function $h(y)$ is an arbitrary weighting function (see
Tremaine \& Weinberg \markcite{tm84}1984). Although the method was intended to
measure bar pattern speeds, it can equally be applied to any strong and fairly
open pattern, such as spiral arms.

The continuity assumption is valid for dust-free old stellar populations, but
may well break down in dusty systems, or when using gas as a tracer
(possibility of ionisation, molecular conversion, optical thickness, etc.). In
fact, while Tremaine \& Weinberg \markcite{tm84}(1984) succeeded in applying
their method to an artificial dataset generated from an $N$-body simulation,
they failed on \ion{H}{1} data from the barred spiral galaxy NGC~5383
(Sancisi, Allen, \& Sullivan \markcite{sas79}1979). The method is fairly
sensitive to noise because only the non-symmetric (i.e.\ odd in $x$) part of
$\Sigma$ contributes to the integrals in equation~(\ref{eq:tm}).

Kent \markcite{k87}(1987) applied the TW method to the barred spiral galaxy
NGC~936. The stellar component of NGC~936 is ideal for this application
because it has a strong bar, high surface brightness, high rotation velocity,
and is devoid of dust and star formation. The orientation of the system is
also favourable. Kent carefully estimated the sources of error in measuring
$\Omega_p$. Unfortunately, the relatively poor quality of the kinematical data
(by today's standards) led to inconclusive results. He derived an average
pattern speed of $\Omega_p=8.4\pm2.9$~km~s$^{-1}$~arcsec$^{-1}$
($\Omega_p=104\pm36$~km~s$^{-1}$~kpc$^{-1}$ for $D=16.6$~Mpc). With Kormendy's
\markcite{k84}(1984) rotation curve, corrected for non-circular motions, and a
deprojected bar length of 52\arcsec, corotation then lies between 0.54 and
1.21 bar semi-lengths. This is only marginally consistent with the general
result from orbital calculations and self-consistent models of barred spiral
galaxies, that corotation must occur outside the end of the bar and is
generally located just beyond its ends (see e.g.\ Sellwood
\markcite{se81}1981; Teuben \& Sanders \markcite{ts95}1985; Athanassoula
\markcite{a92b}1992b).

Merrifield \& Kuijken \markcite{mk95}(1995; see also Kuijken \& Merrifield
\markcite{km96}1996) reapplied the TW method to NGC~936, this time with better
data and methods. They did not use equation~(\ref{eq:tm}) directly, but rather
used the equivalent formula
\begin{equation}
\label{eq:mk}
\Omega_p\sin i=\frac{\langle \bar{v}_{los}\rangle-v_{sys}}{\langle x\rangle}
\;\;\;\;\mbox{(for a given $y$)},
\end{equation}
where the brackets represent luminosity-weighted averages along the slit. The
advantage of this procedure is that the long-slit data can be co-added along
the spatial axis to measure $\langle\bar{v}_{los}\rangle$, and along the
dispersion axis to measure $\langle x\rangle$, yielding higher signal-to-noise
measurements. To obtain an unbiased measure of $\langle\bar{v}_{los}\rangle$,
Merrifield \& Kuijken \markcite{mk95}(1995) also used a deconvolution
algorithm allowing for asymmetric Doppler broadening of the spectral lines.
They obtained a well-defined average pattern speed of
$\Omega_p=4.8\pm1.1$~km~s$^{-1}$~arcsec$^{-1}$
($\Omega_p=60\pm14$~km~s$^{-1}$~kpc$^{-1}$ for $D=16.6$~Mpc). This value puts
corotation at $69\pm15\arcsec$, or $1.3\pm0.3$ bar radii, in good agreement
with theoretical expectations.

Using the same procedure, Gerssen, Kuijken, \& Merrifield
\markcite{gkm99}(1999) recently measured the bar pattern speed in the SBa
galaxy NGC~4596, which is similar to NGC~936. They derive a pattern speed of
$\Omega_p=3.9\pm1.0$~km~s$^{-1}$~arcsec$^{-1}$
($\Omega_p=52\pm13$~km~s$^{-1}$~kpc$^{-1}$ for $D=15.7$~Mpc). This again
indicates a fast bar, placing corotation just outside the end of the bar, at
approximately 1.25 bar semi-lengths (see also Kent \markcite{k90}1990).
\section{NGC~2915\label{sec:ngc2915}}
\nopagebreak
\subsection{Background\label{sec:background}}
\nopagebreak
\placetable{ta:basic}

NGC~2915 is a nearby relatively isolated BCD galaxy. \markcite{mmc94}MMC94
carried out optical imaging and spectroscopy of NGC~2915 to study its
structure, stellar populations, and star formation history.
\markcite{mcbf96}MCBF96 obtained \ion{H}{1} synthesis data to study the
neutral hydrogen distribution and kinematics, and NGC~2915's overall mass
distribution (including dark matter). Table~\ref{ta:basic} summarises the
basic properties of NGC~2915.

The photometry of \markcite{mmc94}MMC94 shows that NGC~2915 has two distinct
stellar populations: (1) an exponential red diffuse population in the outer
parts ($r\geq35\arcsec=0.90$~kpc), with a scale length
$\alpha_B^{-1}=25\farcs6$ (660~pc), an extrapolated central surface brightness
$B(0)_c=22.44$~mag~arcsec$^{-2}$ (corrected for inclination and extinction),
and constant color; and (2) a slightly peaked core population ($r<35\arcsec$),
with increasingly bluer color toward the center. The diffuse population is
similar to that of dwarf elliptical galaxies, but the core is lumpy, contains
a young stellar population, and is the locus of high mass star formation.
Spectroscopy in the core shows strong Balmer and \ion{Ca}{2} H+K absorption
lines, and strong narrow emission lines with ratios typical of high excitation
low metallicity \ion{H}{2} regions. A velocity difference of 60~km~s$^{-1}$ is
detected across the inner galaxy. If this gradient is due solely to rotation,
it corresponds to a deprojected rotational velocity of
$V_r=50\pm21$~km~s$^{-1}$ at $r=32\arcsec$ (neglecting the effects of the
bubbles seen in H$\alpha$ by Marlowe et al.\ \markcite{mhws95}(1995)). The
associated mass $M_{\mbox{\scriptsize
Core}}=4.8\times10^8\;M_{\mbox{\scriptsize \sun}}$ for the optical core,
with a mass-to-light ratio $M_{\mbox{\scriptsize
Core}}/L_B=1.3\;M_{\mbox{\scriptsize \sun}}/L_{B,\mbox{\scriptsize
\sun}}$.
\placetable{ta:curves}
\placefigure{fig:h1+optical}
\placefigure{fig:velocity+optical}

The neutral hydrogen observations of NGC~2915 (\markcite{mcbf96}MCBF96) reveal
an \ion{H}{1} distribution very different from the optical.
Figure~\ref{fig:h1+optical} shows the naturally weighted total \ion{H}{1}
intensity map overlaid on an optical image of NGC~2915. The \ion{H}{1}
distribution is unusually extended, up to 5~$D_{\mbox{\scriptsize Ho},B}$
(22.6~$\alpha_B^{-1}$), and it has a striking late-type barred spiral galaxy
morphology, with two extended spiral arms starting at the end of a central
bar. This is unusual as bars are generally \ion{H}{1} poor. At higher
resolution, the \ion{H}{1} bar is resolved into two clouds; its semi-length is
about $2\farcm4$ (3.7~kpc), and it corresponds roughly in position,
orientation, and size to the optical emission. The \ion{H}{1} surface density
profile is tabulated in Table~\ref{ta:curves}. The total \ion{H}{1} mass is
$M_{\mbox{\scriptsize HI}}=9.6\times10^8\;M_{\mbox{\scriptsize \sun}}$, while
the \ion{H}{1} mass in the bar is only about
$5-7.5\times10^7\;M_{\mbox{\scriptsize \sun}}$.
Figure~\ref{fig:velocity+optical} shows the \markcite{mcbf96}MCBF96 naturally
weighted \ion{H}{1} velocity field of NGC~2915, again overlaid on an optical
image. It clearly shows the pattern of a rotating disk (with an oval
distortion), and suggests the presence of a warp in the outer parts.
\markcite{mcbf96}MCBF96 derived a rotation curve for NGC~2915 using the
standard tilted ring algorithm ROTCUR (Begeman \markcite{b89}1989). The
position angle varies by about 15\arcdeg\ and the inclination by about
10\arcdeg\ outside of the barred region (where the effects of the bar and a
possible warp are hard to disentangle), supporting the presence of a warp in
the outer disk. The \ion{H}{1} velocity dispersion outside the bar region is
the usual $\sigma_v\approx$~8~km~s$^{-1}$, but it increases up to
40~km~s$^{-1}$ in the central regions (see Table~\ref{ta:curves}), a value
comparable to the rotation velocities obtained from ROTCUR. This indicates
that pressure support is important in the center. \markcite{mcbf96}MCBF96
applied an asymmetric drift correction to the rotation curve to produce the
``circular velocity curve'' needed for mass modelling. The circular velocity
is tabulated in Table~\ref{ta:curves}. It increases rapidly with radius in the
inner parts, reaching 80~km~s$^{-1}$ by $r=150\arcsec$ (3.9~kpc), stays flat
until $r=400-450\arcsec$ (10.3-11.6~kpc), and increases again up to
$\approx$90~km~s$^{-1}$ at the last measured point
($r\approx600\arcsec$~=~15.4~kpc).

\markcite{mcbf96}MCBF96 constructed three-component mass models to interpret
the circular velocity curve. The models were composed of: (1) a stellar disk,
obtained from the surface brightness profiles of \markcite{mmc94}MMC94 (the
mass-to-light ratio $M/L_B$ is a free parameter); (2) an \ion{H}{1} disk,
based on their observations; and (3) a spherical dark halo (two density
distributions were used, both with a central density $\rho_0$ and core radius
$r_c$ as free parameters). All models are dominated by dark matter at all
radii. In fact, NGC~2915 is one of the ``darkest'' disk galaxies known. Its
dark matter core is very dense and compact, which may prove crucial to
understand its pattern speed (see \S~\ref{sec:selfcon}).
\markcite{mcbf96}MCBF96's best fit model\footnote{Their model D, with a dark
matter density distribution $\rho(r)=\rho_0[1+(r/r_c)^2]^{-1}$.} has
$M/L_B=1.2\;M_{\mbox{\scriptsize \sun}}/L_{B,\mbox{\scriptsize \sun}}$,
$\rho_0=0.10\pm0.02\;M_{\mbox{\scriptsize \sun}}$~pc$^{-3}$, and
$r_c=1.23\pm0.15$~kpc, yielding a total mass
$M_T=2.7\times10^{10}\;M_{\mbox{\scriptsize \sun}}$,
$M_T/L_B=76\;M_{\mbox{\scriptsize \sun}}/L_{B,\mbox{\scriptsize \sun}}$, and
$M_{dark}/M_{luminous}=19$ within the last measured point.
\subsection{Pattern Speed\label{sec:patspeed}}
\nopagebreak
We now attempt to measure the pattern speed $\Omega_p$ of the figure in the
disk of NGC~2915 using the \ion{H}{1} data of \markcite{mcbf96}MCBF96. As
mentioned in \S~\ref{sec:direct}, the continuity equation imposes a strong
constraint on the applicability of the TW method, and gaseous systems are
usually thought to be poor candidates for a measurement. NGC~2915 is an
unusual galaxy, however, and we argue that the TW method should work for its
\ion{H}{1} distribution. Firstly, because the stellar component of NGC~2915 is
confined to the inner regions of its \ion{H}{1} distribution, and the
star-forming core to the very center of the \ion{H}{1} bar, there is no source
of ionising photons over most of the \ion{H}{1} disk (except for the weak
metagalactic radiation field, and possibly some radiation from the distant
core reaching the warped regions). Meurer et al.\ (\markcite{mfbka99}1999)
searched for \ion{H}{2} regions in the outer disk, but found none. This
situation is in contrast to that in
NGC~5383, where Tremaine \& Weinberg (\markcite{tw84}1984) failed to obtain a
reliable measurement of the pattern speed using neutral hydrogen observations.
Furthermore, the \ion{H}{1} column density is low enough in most of the disk
(where the velocity dispersion $\sigma_v\approx$~8~km~s$^{-1}$) that optical
thickness problems should not affect the observed \ion{H}{1} intensity. Only
in the inner parts of the bar, where the optical disk lies, does the column
density reach $N_{HI}=10^{21}$~cm$^{-2}$, but there the \ion{H}{1} velocity
dispersion is larger ($\sigma_v\approx$~20-40~km~s$^{-1}$), so again the
optical thickness is expected to be low. No molecular data on NGC~2915 are
available in the litterature, so we do not have any information on possible
\ion{H}{1}-molecule conversion at this point. However, the low \ion{H}{1}
column density again suggests that this effect is not important. In order to
properly constrain its molecular content, we are nevertheless pursuing
molecular line observations of NGC~2915. For all these
reasons, and because the bar and spiral arm pattern is clear and strong, we
believe that the TW method applied to the \ion{H}{1} synthesis data of
NGC~2915 should give a realistic measure of its pattern speed. We note that
the orientation of the system is ideal, with the disk at intermediate
inclination and the bar about 30\arcdeg\ from the major axis of the disk.
\placefigure{fig:h1+tm}
\placefigure{fig:tmresults}

We apply the TW method to NGC~2915 using \markcite{mcbf96}MCBF96's naturally
weighted total \ion{H}{1} intensity map and velocity field (see
Fig.~\ref{fig:h1+optical} and \ref{fig:velocity+optical}). Additional
structural and kinematical parameters are required to use the TW method (see
\S~\ref{sec:direct}), but these are all provided by the tilted-ring analysis
of \markcite{mcbf96}MCBF96. We use averages of their parameters for the outer
disk, outside of the barred region ($r>150\arcsec$): the center of the disk
$(x_c,y_c)=(0\arcsec,0\arcsec)$ (with respect to the pointing center of
$\alpha=9^{\mbox{\rm h}}26^{\mbox{\rm m}}11^{\mbox{\rm s}}$,
$\delta=-76\deg37\arcmin35\arcsec$), $v_{sys}=467$~km~s$^{-1}$, and the
position angle of the line of nodes PA~=~117\arcdeg. We use
equation~(\ref{eq:tm}) directly, without a weighting function $h(y)$, and for
all possible positions $y$ (equivalent to using $h(y)=\delta(y-y_o)$ for each
$y_o$). Because the major axis traces the position of the line of nodes, $y$
is effectively the position of the integration axis along the minor axis of
the galaxy. Each offset $y$ provides an independent measurement of the pattern
speed. The coordinate system is shown in Figure~\ref{fig:h1+tm}, and
Figure~\ref{fig:tmresults}a shows the derived pattern speed as a function of
the offset $y$. The measurements are roughly consistent with each other for
most offsets. Only for a few isolated points and for offsets between 0 and
200\arcsec\ on the NW side of the galaxy (positive offsets) are the values of
$\Omega_p\sin i$ significantly discrepant (but see below). We recall that the
TW method assumes a unique pattern speed for the whole disk, so all offsets
should yield similar pattern speeds.

The discrepant region is associated with the northernmost of the two clouds
identified by \markcite{mcbf96}MCBF96, located at the northern end of the bar.
Around that region, both the total \ion{H}{1} intensity distribution and the
velocity field show a strong deviation from their largescale pattern. The
discrepancies in the derived pattern speeds probably arise from the fact that
the cloud breaks the pattern in the density distribution and/or the velocity
field, an effect to which equation~(\ref{eq:tm}) is very sensitive. Away from
this problematic region, the pattern speed is well-defined: the entire SE half
of the galaxy and the farthest regions in the NW half both yield fairly
consistent values for $\Omega_p\sin i$. We thus believe that the TW method
works using the \ion{H}{1} component of NGC~2915 and gives a reliable estimate
its the pattern speed.

In our calculations, we assumed that the parameters required to use the TW
method were fixed over the whole disk. The center of the disk is well-defined,
and only the x-component of the center ($x_c$) affects the calculations
anyway. The systemic velocity is also well constrained, with a small
uncertainty of a few km~s$^{-1}$. However, as shown by the tilted-ring
analysis of \markcite{mcbf96}MCBF96, both the inclination $i$ and position
angle PA vary significantly with radius in NGC~2915. A change in inclination
only affects the pattern speed measurements through the deprojection, but a
change in position angle materially affects the calculations themselves. We
have therefore reapplied the TW method to the \ion{H}{1} data using a range of
values for $x_c$, $v_{sys}$, and PA. Although no combination of parameters
yields perfectly equal pattern speeds for all offsets, most cases with
plausible values show a regular behaviour, with only a restricted range of
offsets having very discrepant measurements.

A typical good case is presented in Figure~\ref{fig:tmresults}b. For all
combinations of parameters, the flat portions of the $\Omega_p\sin i$
``distribution'' always lie between $\Omega_p\sin
i=0.12$~km~s$^{-1}$~arcsec$^{-1}$ and $\Omega_p\sin
i=0.22$~km~s$^{-1}$~arcsec$^{-1}$. We therefore adopt $\Omega_p\sin
i=0.17\pm0.05$~km~s$^{-1}$~arcsec$^{-1}$ ($\Omega_p\sin
i=6.6\pm1.9$~km~s$^{-1}$~kpc$^{-1}$ for $D=5.3$~Mpc) as our estimate of the
pattern speed in NGC~2915. The uncertainty given represents half the range of
measured values for $\Omega_p\sin i$, and is not a standard error estimate. It
is clear from our measurements that formal errors are much smaller than the
errors introduced by the uncertainties in the structural and kinematical
parameters of the disk, or the errors introduced by the fact that we are
working with a less-than-perfect system. From the outer parts of the disk
again, we get an average inclination $i=56\pm3\arcdeg$. This gives a
deprojected pattern speed $\Omega_p=0.21\pm0.06$~km~s$^{-1}$~arcsec$^{-1}$
($8.0\pm2.4$~km~s$^{-1}$~kpc$^{-1}$).

The 30\% uncertainty in $\Omega_p$ is comparable to that achieved by
Merrifield \& Kuijken \markcite{mk95}(1995) for NGC~936, but is still large
given that we are using gas kinematics with the advantage of complete
two-dimensional spatial coverage. On the other hand, our error estimate (half
the range of measured values) is very conservative. As suggested by Tremaine
\& Weinberg \markcite{tm84}(1984), we tried using odd weighting functions
(e.g.\ $h(y)=\delta(y-y_0)-\delta(y+y_0)$; $y_c$ now affecting the
calculations directly). Such weighting eliminates the contribution of
perturbations with odd azimuthal wavenumbers to the integrals in
equation~(\ref{eq:tm}), and diminishes the effect of centering errors.
However, it did not improve our results, because discrepant regions on one
side of the minor axis contaminate the combined measurements from both sides.
Although we cannot exclude the possibility that the continuity equation is not
fully satisfied for the \ion{H}{1} in some parts of NGC~2915 (particularly the
central regions), the main sources of uncertainty in our measurements are the
poorly constrained disk parameters (mainly PA), in turn related to the
presence of a strong warp in the outer disk. Tremaine \& Weinberg
\markcite{tm84}(1984) note that a warp may affect the pattern speed
determination, acting as a non-axisymmetric perturbation with zero pattern
speed. There may be some short-lived features in the disk (such as the
structure seen near the NW cloud), but their effect is hard to quantify. We
repeated the calculations using only the inner barred region of the galaxy,
but the results were inconclusive. It is far from clear that the method should
work in such a case anyway, as the integrals in eq.~[\ref{eq:tm}] should
extend over the whole system.
\section{The Nature of the Pattern\label{sec:natpat}}
\nopagebreak
\subsection{Self-Consistent Bars\label{sec:selfcon}}
\nopagebreak
Orbital calculations in barred disk potentials show that bars should end at or
inside their corotation radius (see, e.g., Contopoulos \markcite{c80}1980;
Athanassoula \markcite{a92a}1992a; for a review see Contopoulos \& Grosb\o l
\markcite{cg89}1989). The argument is that the bar-shaped $x_1$ orbits,
believed to support self-consistent bars, extend close to but not past
corotation.

$N$-body studies show that rotationally supported stellar disks are globally
unstable to bi-symmetric distortions and quickly form strong bars. This is
seen in both two-dimensional (e.g.\ Sellwood \markcite{se81}1981; Athanassoula
\& Sellwood \markcite{as86}1986) and three-dimensional (e.g.\ Combes \&
Sanders \markcite{cs81}1981; Raha et al.\ \markcite{rsjk91}1991) simulations.
These instabilities can be identified with the dominant linear modes found in
analytical global stability analyses of axisymmetric disks (e.g.\ Kalnajs
\markcite{k71}1971, \markcite{k77}1977). The main characteristic of the
stellar bars formed in such simulations is that the they are fast, generally
ending at or just inside corotation (where $\Omega_p=\Omega(r)$).

This property of self-consistent barred disks does not appear to hold in
NGC~2915; the gaseous bar ends well within corotation. We take the ends of the
bar to be where the density of the rectangular-shaped central component of the
\ion{H}{1} distribution drops rapidly. This is also the region where the
spiral arms start. The observed semi-length of the bar is then about
145\arcsec, with a deprojected value of $r_b\approx180\arcsec$. Our adopted
pattern speed $\Omega_p=0.21\pm0.06$~km~s$^{-1}$~arcsec$^{-1}$ (see
\S~\ref{sec:patspeed}) and \markcite{mcbf96}MCBF96's circular velocity curve
(see Table~\ref{ta:curves}) put corotation at a radius
$r_{co}=390^{+\infty}_{-80}$ arcsec (see Fig.~\ref{fig:resonances}). The ratio
of corotation to the bar semi-length is thus very large, $r_{co}/r_b>1.7$,
indicating a {\em slowly} rotating bar.

Such a slow pattern speed is inconsistent with the the simulations mentioned
above. However, all these studies used models where luminous mass dominates
the inner (optical) regions of the galaxies. \markcite{mcbf96}MCBF96 showed
that this is not the case in NGC~2915, where dark matter dominates at all
radii. In fact, the dense and compact dark matter core of NGC~2915 may be the
clue to its slow pattern speed. Dense dark halos have been predicted to slow
down bars significantly, as the bar can transfer angular momentum to the outer
disk and halo efficiently through dynamical friction (see Sellwood
\markcite{s80}1980; Weinberg \markcite{w85}1985; Debattista \& Sellwood
\markcite{ds98}1998). Because dak matter is generally more important in
late-type spirals, where it is difficult to measure pattern speeds, present
support for these models is marginal (see, e.g., Elmegreen
\markcite{e96}1996). NGC~2915 may thus represent the strongest observational
evidence so far for bar deceleration by dark halos with high central
densities. The mass model of \markcite{mcbf96}MCBF96, the $N$-body simulations
of Debattista \& Sellwood \markcite{ds98}(1998), and the bar pattern speed
measured here are all consistent with each other. However, the effects of
gaseous processes on bar deceleration remain to be fully investigated. These
have not been taken into account in the aforementioned studies, but are
certainly relevant in the case of NGC~2915. Gas inflow, in particular, could
lead to a speed up of the bar.
\subsection{Bar or Spiral Arms Pattern Speed?\label{sec:barspi}}
\nopagebreak
The TW method measures the pattern speed of a disk, assumed to be unique. While
it was designed to measure bar pattern speeds, it equally applies to any
pattern present in the disk, as long as it is strong enough. It is therefore
important to establish which component's pattern speed we are measuring. Are
we measuring the pattern speed of the bar, the spiral arms, or a common
pattern speed for both?

Sellwood \& Sparke \markcite{ss88}(1988) suggest that spiral arms in barred
spiral galaxies could have a different (lower) pattern speed than the bar.
Their work was aimed primarily at explaining the presence of dust lanes on the
inner edge of spiral arms in many objects. A lower pattern speed for the
spiral pattern would allow the gas to overtake that pattern and therefore lie
within its corotation radius in that region, while still retaining the view
that bars end just inside {\em their} corotation radius. Sellwood \& Sparke
\markcite{ss88}(1988) suggestion is supported by $N$-body simulations of
barred galaxies showing outer spiral patterns slower than the bars (e.g.\ 
Sparke \& Sellwood \markcite{ss87}1987).

Tagger et al.\ \markcite{tsap87}(1987) and Sygnet et al.\ 
\markcite{stap88}(1988) develop the idea that the non-linear coupling of modes
with different pattern speeds could sustain spiral structures for a long time.
For the mechanism to be efficient, resonances of the two patterns must be
closely spaced. They observe this effect in many $N$-body simulations, where
the corotation radius of the bar and the inner Lindblad resonance (ILR) of the
spiral pattern are very close (this is also the region where the bar ends and
the spiral arms begin). The beat wave generated is then axisymmetric and close
to its Lindblad resonance and it couples non-linearly to the bar and spiral
arms.

If we assume that the pattern speed measured here is that of the spiral arms,
and that the corotation radius of the bar lies close to the ILR of the spiral
pattern, as suggested by Tagger et al.\ \markcite{tsap87}(1987) and Sygnet et
al.\ \markcite{stap88}(1988), we obtain a bar corotation radius
$r_{co}<50\arcsec$ (see Fig.~\ref{fig:resonances}), well inside the bar
observed in \ion{H}{1} ($r_b\approx180\arcsec$). Such a view would be
difficult to defend, considering the body of analytical and numerical work
which requires the corotation radius to lie at or beyond the end of the
bar. This is a strong indication that we are in fact measuring the pattern
speed of the bar.

Furthermore, considering the form of equation~(\ref{eq:tm}), we see that for
small minor axis offsets ($y<90\arcsec$) both the central (barred) and
external (non-barred) regions of the galaxy contribute to the integrals, while
for large offsets ($y>90\arcsec$), the integration axis does not cross the
barred region and only the outer parts of the disk contribute to the pattern
speed calculations. If the bar and spiral arms had significantly different
pattern speeds, we would expect the TW method to fail for small offsets.
However, excluding the discrepant points 0-200\arcsec\ on the NW side of the
galaxy (which we attribute to a large \ion{H}{1} cloud, see
\S~\ref{sec:patspeed}), the pattern speed measurements yield a relatively
constant value for all offsets and for most reasonable combinations of
parameters.  Considering the relatively small extent of the barred region and
the importance of the flux in the outer parts of the disk, it might appear
that the spiral arms dominate the contribution to the integrals in
equation~(\ref{eq:tm}) even in the central regions. This is not so, however.
For small offsets, both the inner barred region and the outer parts of the
disk contribute significantly to the numerator. It is therefore very likely
that a unique pattern speed is being measured. This view is further
strengthened by the fact that the spiral arms in NGC~2915 appear to start from
the end of the bar (although Sellwood \& Sparke \markcite{ss88}(1988) showed
that this is not a strong argument).

We therefore adopt the view that the pattern speed which we have measured in
\S~\ref{sec:patspeed} {\em is} the unique, common pattern speed of both the
bar and the spiral arms, although we cannot entirely exclude the possibility
that it is the pattern speed of the spiral arms alone.
\section{The Origin of the Pattern\label{sec:oripat}}
\nopagebreak
Many theories exist to explain the formation of bar and spiral patterns in
disk galaxies. We will now focus on some of these. We note that the issue of
the origin of the pattern in NGC~2915 is largely indepedent of the measured
pattern speed. For example, is the pattern a phenomenon of the disk itself, or
is it driven by other effects? We will show that the existence of a pattern in
NGC~2915 is somewhat problematic.
\subsection{Gravitational Interactions\label{sec:interactions}}
\nopagebreak
Tidal triggering is a potentially important way to form bars, especially in
dense environments. In the case of parabolic prograde planar encounters,
Noguchi \markcite{n87}(1987) finds that self-consistent disks (with a static
halo) develop strong spiral structures over their entire length, with a
bar-like pattern in the inner regions. Gerin, Combes, \& Athanassoula
\markcite{gca90}(1990) show that interaction with a companion accelerates bar
formation. The sense of interaction (prograde or retrograde) affects the bar
growth rate and shape, with lesser effects on the strength and pattern speed
of the resulting bar. When a bar already exists, its strength and pattern
speed can again be affected by an interaction.

It seems unlikely that such interaction scenarios can explain the structure
seen in the \ion{H}{1} disk of NGC~2915. No evidence of interaction is seen in
the \ion{H}{1} data of \markcite{mcbf96}MCBF96, and the only catalogued
potential interaction partner for NGC~2915 within 5\arcdeg\ is a low
luminosity, low surface brightness object, SGC~0938.1-7623, at a radius of
42\arcmin\ (projected distance of 65~kpc at the distance of NGC~2915). This
object does not have a catalogued optical or \ion{H}{1} velocity however, so
although there are no indications that NGC~2915 is interacting, we cannot
definitively exclude this possibility.
\subsection{Swing Amplification\label{sec:swing}}
\nopagebreak
\placefigure{fig:Xr}

Swing amplification (Toomre \markcite{t81}1981) can generate spiral arms in a
disk as transient tidal features. Although Toomre \markcite{t81}(1981)
considered a stellar disk, his treatment should equally apply to gaseous
disks. Swing amplification is a local cooperative effect: shear, epicyclic
motions, and self-gravity all contribute to support the spiral pattern,
originally excited by a small disturbance. In short, the epicyclic motions
resonate with the shear flow as the spiral pattern wraps, and the spiral arms
are sustained. The self-gravity of the arms then leads to a gravitational
instability (Goldreich \& Lynden-Bell \markcite{gl65}1965). The effect can be
summarised (somewhat simplified by neglecting the shear) in Toomre's parameter
$X$, given by
\begin{equation}
\label{eq:X}
X=\frac{\lambda}{\lambda_{\mbox{crit}}},
\end{equation}
where
\begin{equation}
\label{eq:lambdacrit}
\lambda_{\mbox{crit}}=\frac{4\pi^2{\mbox{G}}\mu}{\kappa^2},
\end{equation}
$\mu$ is the surface density, and $\kappa$ is the epicyclic frequency. Swing
amplification is efficient for $X\lesssim3$ (taking into account the effect of
shear and random motions; see, e.g., Fig.~7 of Toomre \markcite{t81}1981). For
an extended disk, we can write $\lambda=2\pi r/m$, where $m$ is the azimuthal
wavenumber of the spiral pattern considered, and
\begin{equation}
\label{eq:Xr}
X(r)=\frac{r\kappa(r)^2}{2\pi\mbox{G}m\mu(r)}
\end{equation}
(see Athanassoula, Bosma, \& Papaioannou \markcite{abp87}1987). 

Figure~\ref{fig:Xr} shows $X(r)$ for NGC~2915 for two and four-armed spiral
patterns. We use here the surface density of the luminous matter only (gaseous
and stellar), assuming that the dark matter is not responsive to the density
perturbation created by the spiral pattern (Toomre \markcite{t81}1981
considered a fixed halo). In Figure~\ref{fig:Xr}a, only the \ion{H}{1}
distribution from \markcite{mcbf96}MCBF96 is included (uncorrected for
inclination; see Table~\ref{ta:curves}). It is clear from the figure that
$X(r)$ is nowhere near a value of 3. In fact, even for $m=4$, $X(r)\gtrsim10$
for almost all radii. In Figure~\ref{fig:Xr}b, the stellar content is also
included. To construct the surface density profile, we used the $B$ surface
brightness profile of \markcite{mmc94}MMC94 for $r<125\arcsec$, and an
extrapolation of their exponential fit to the outer parts of the disk for
larger radii: $B_c=21.79$~mag~arcsec$^{-2}$ (corrected for extinction but not
inclination) and $\alpha_B^{-1}= 25\farcs6$, with
$M/L_B=1.2\;M_{\mbox{\scriptsize \sun}}/L_{B,\mbox{\scriptsize \sun}}$ (as in
the adopted mass model of \markcite{mcbf96}MCBF96). The resulting surface
brightness profile is tabulated in Table~\ref{ta:curves}. Even with the
stellar component, $X(r)\gtrsim10$ for most radii (any correction for
inclination would further decrease the importance of luminous matter). Small
values of $X(r)$ are present in the inner regions only: $X(r)\leq5$ for
$r<75\arcsec$ only. We note however that those regions are within the rapidly
rising portion of the rotation curve, where the bar resides and the shear is
low, while swing amplification relies on shear to amplify the spiral pattern.

The work by Toomre \markcite{t81}(1981) refers to a flat rotation curve,
$V_c(r)=\mbox{constant}$, which is not exactly the case here. The
amplification criterion $X\lesssim3$ increases for higher shear rates (e.g.\ 
$X\lesssim6$ for a Keplerian rotation curve, $V_c(r)\propto r^{-1/2}$) and
decreases for smaller shear (e.g.\ $X\lesssim1.5$ for $V_c(r)\propto r^{1/2}$;
Toomre, private communication). So, if anything, the constraints on swing
amplification are even stronger in the case of NGC~2915, especially in the
inner parts.
\placefigure{fig:Qr}

Toomre's \markcite{t64}(1964) stability parameter $Q$ also affects the
efficiency of the swing amplifier mechanism. The stellar disk of NGC~2915 does
not extend into the spiral pattern region, so we consider only the gaseous
disk. Then,
\begin{equation}
\label{eq:Qr}
Q(r)=\frac{v_s(r)\kappa(r)}{\pi{\mbox{G}}\mu(r)},
\end{equation}
where $v_s$ is the sound speed of the gas, taken here as the velocity
dispersion. As $Q$ increases, the maximum swing amplification factor
decreases. For $V_c(r)=\mbox{constant}$, $Q\gtrsim2$ curbs most of the
amplification (see, again, Fig.~7 of Toomre \markcite{t81}1981). The criterion
becomes $Q\gtrsim3$ for $V_c(r)\propto r^{-1/2}$ and $Q\gtrsim1.5$ for
$V_c(r)\propto r^{1/2}$ (Toomre, private communication). Figure~\ref{fig:Qr}
shows $Q(r)$ for NGC~2915 using both a constant velocity dispersion, equal to
the velocity dispersion in the outer parts of the \ion{H}{1} disk, and the
\ion{H}{1} velocity dispersion profile of \markcite{mcbf96}MCBF96, tabulated
in Table~\ref{ta:curves}. We see that $Q(r)\gtrsim5$ everywhere, so the
amplification factor would be small. In particular, $Q$ is high in the inner
regions. The sound speed would have to be less than 2-3~km~s$^{-1}$ to make
$Q<2$ over a significant range of radii. Even then, $X$ would still be too
high for the swing amplifier to work efficiently.

Given the behaviour of $X(r)$ and $Q(r)$, and if the analogy between Toomre's
\markcite{t81}(1981) stellar treatment and the gaseous treatment adopted here
is at least qualitatively valid\footnote{As mentioned by Toomre
  \markcite{t81}(1981) based on the work of Bardeen \markcite{b75}(1975) and
  Aoki, Noguchi, \& Iye \markcite{ani79}(1979), gaseous disks are probably
  more stable than stellar ones, which would strengthen our conclusions.}, it
is clear that swing amplification cannot explain the spiral pattern seen in
the disk of NGC~2915. We note that both $X$ and $Q$ scale inversely with the
distance adopted, so a large error in the distance could weaken or invalidate
our conclusion (if the real distance is larger). However, the error estimate
of \markcite{mmc94}MMC94 does not allow this (they quote $D=5.3\pm1.6$~Mpc). A
distance of at least 10--15~Mpc would be required to make swing amplification
efficient. Furthermore, given the very extended mass distribution of NGC~2915,
it is unlikely that the observed spiral pattern results from an edge-mode, as
discussed by Toomre \markcite{t81}(1981).

At this point, we conclude that the bar and spiral patterns in NGC~2915 are
unlikely to be caused by gravitational interaction or swing amplification. We
must therefore search for an alternative formation mechanism.
\subsection{Resonance Ring Formation\label{sec:resring}}
\nopagebreak
\placefigure{fig:resonances}

The arms of spiral galaxies frequently wrap to form ring-like structures. The
theory of resonance ring formation (Schwarz \markcite{s81}1981,
\markcite{s84}1984; Byrd et al.\ \markcite{brsbc94}1994) relates the shapes
and positions of the rings to the presence of resonances in the disks. An
examination of the total \ion{H}{1} map of NGC~2915 suggests the presence of
two such rings. For a given rotation curve, the bar pattern speed determines
the existence and position of the resonances, and therefore the kind and
extent of the orbit families present in the disk (Sellwood \& Wilkinson
\markcite{sw93}1993). We can thus use resonance ring theory to see if our
pattern speed measurement is consistent with the location of the
(pseudo-)rings seen in NGC~2915, and determine whether the spiral arms are
driven by the bar.

Using \markcite{mcbf96}MCBF96 circular velocity curve, we can calculate the
angular frequency of circular motion $\Omega$ and the associated Lindblad
precession frequencies $\Omega+\kappa/2$, $\Omega-\kappa/2$, and
$\Omega-\kappa/4$ (see Table~\ref{ta:curves}). These are shown in
Figure~\ref{fig:resonances}, along with the measured pattern speed and the
position of the pseudo-rings in the disk of NGC~2915.

The innermost points in Figure~\ref{fig:resonances} are affected by beam
smoothing but suggest the presence of an ILR ($\Omega_p=\Omega-\kappa/2$) at
$r<50\arcsec$. Nuclear rings are believed to be associated with ILRs, and
usually form easily recognisable annuli of \ion{H}{2} regions in the inner
parts of galaxies. The shallower the rotation curve in the inner parts, the
larger the nuclear ring formed (Byrd et al.\ \markcite{brsbc94}1994). However,
no such ring is visible in the photometry of \markcite{mmc94}MMC94 or in the
Fabry-Perot H$\alpha$ images of Marlowe et al.\ \markcite{mhws95}(1995). The
resolution of \markcite{mcbf96}MCBF96 \ion{H}{1} data is too low to resolve a
ring of radius smaller than about 1\farcm5.

The inner spiral arms seen in the total \ion{H}{1} intensity map of NGC~2915
(Fig.~\ref{fig:h1+tm}) appear to form a pseudo-ring just outside the end of
the bar, at a deprojected radius of $215\pm15\arcsec$. Inner rings are usually
associated with the inner second-harmonic resonance
($\Omega_p=\Omega-\kappa/4$) which, in this case, occurs between about
190\arcsec\ and 340\arcsec. This is just consistent with the position of the
inner pseudo-ring.

The outer spiral arms seen at the edge of the \ion{H}{1} disk in
Figure~\ref{fig:h1+tm} form an outer pseudo-ring at $r=400\pm15\arcsec$. Outer
rings are associated with the outer Lindblad resonance
($\Omega_p=\Omega+\kappa/2$). However, the $\Omega+\kappa/2$ curve in
Figure~\ref{fig:resonances} does not intersect the range of measured values
for $\Omega_p$, which suggests that no outer Lindblad resonance is present in
the disk of NGC~2915. Therefore, no outer ring would be expected.

Resonance ring theory has only limited success in explaining the position of
the various (pseudo-)rings in NGC~2915 (nuclear, inner, and outer rings). It
could be that the pattern in NGC~2915 is still evolving, and that the spiral
arms are still in the process of forming rings. This seems unlikely, however,
since no nuclear ring is detected, and the timescale for the formation of
nuclear rings is very short ($10^7$~-- $10^8$~yr; Combes \markcite{c93}1993).
Stellar mass loss or gas infall can also delay the formation of rings (Schwarz
\markcite{s81}1981), but these are improbable as most of the \ion{H}{1} disk
is devoid of stars and NGC~2915 is isolated.

Petrou \& Papayannopoulos \markcite{pp86}(1986) proposed a mechanism to
terminate bars well within their corotation radius. This mechanism requires
the presence of a 1:1 resonance ($\Omega_p=\Omega-\kappa$) in the bar,
however, so it is clear from Figure~\ref{fig:resonances} that it is not
relevant in the case of NGC~2915.

We conclude that the structures seen in the disk of NGC~2915 are probably not
due to resonances and that the spiral pattern is not driven by the bar. In
fact, from the positions of the inner and outer pseudo-rings, no single value
of the pattern speed can be derived from resonance ring theory.
\section{Dark Matter and the Structure of NGC~2915\label{sec:dm}}
\nopagebreak
The mass model of \markcite{mcbf96}MCBF96 showed that NGC~2915 is dominated by
dark matter, even in the central (optical) regions of its disk. We now discuss
the possible relations between the properties of this dominant dark matter
distribution and the properties of the \ion{H}{1} pattern.
\subsection{Rotationally Supported Disk Dark Matter\label{sec:diskdm}}
\nopagebreak
Pfenniger, Combes, \& Martinet \markcite{pcm94}(1994; see also Pfenniger \&
Combes \markcite{pc94}1994) propose that the dark matter required dynamically
in disk galaxies is in the form of rotationally supported cold molecular gas.
This idea is attractive for NGC~2915 as it would also provide the extra disk
mass needed to explain the formation of the bar and spiral arms through
gravitational instabilities. Unfortunately, observing such cold gas directly
is difficult (for H$_2$, see Combes \& Pfenniger \markcite{cp97}1997; for CO,
see Lequeux \& Allen \markcite{la93}1993 and Wilson \& Mauersberger
\markcite{wm94}1994).
\placefigure{fig:dmprof}

Figure~\ref{fig:dmprof}a shows how much additional mass is required in the
{\em disk} for the swing amplification mechanism to be efficient ($X=3$
everywhere in the disk; see \S~\ref{sec:swing}). Similarly,
Figure~\ref{fig:dmprof}b shows how much additional disk matter is required to
lower $Q$ to a value of 2 at all radii (the value required to make the disk
gravitationally responsive). Again, those two values scale inversely with
distance (approximately). For comparison, we plot in Figure~\ref{fig:dmprof}c
the observed \ion{H}{1} surface density (Table~\ref{ta:curves}) and the
projected surface density of the dark isothermal {\em sphere} used by
\markcite{mcbf96}MCBF96 to model the circular velocity curve (their model D),
rescaled by a large constant factor (47.7).
\placefigure{fig:muratio}

The similarity between the \ion{H}{1} and the dark matter surface density
profiles seen in Figure~\ref{fig:dmprof}c has been noticed before in other
studies (e.g.\ Bosma \markcite{b78}1978, \markcite{b81}1981; Carignan \&
Beaulieu \markcite{cb89}1989; Broeils \markcite{b92}1992) and has been
interpreted as possible evidence that the \ion{H}{1} and dark matter are
somehow associated. Here, we see also a qualitative similarity between the
observed \ion{H}{1} surface density profile and the surface density profile of
the additional disk matter required to make $Q(r)=2$ (Fig.~\ref{fig:dmprof}b).
It is tempting to regard this as further evidence that the dark matter is
distributed in a disk and follows relatively closely the distribution of
neutral hydrogen. However, the increase in disk surface density required to
make the swing amplifier work efficiently ($X=3$ over most of the disk;
Fig.~\ref{fig:dmprof}a) would lower $Q$ even more, and would make the disk
unstable to axisymmetric modes. Figure~\ref{fig:muratio} shows the ratio of
the surface density obtained by imposing $X(r)=3$ to Kennicutt's
\markcite{k89}(1989) critical surface density for star formation in disks.
This ratio is well above unity for the entire disk, so we would expect to see
evidence of active star formation in the outer \ion{H}{1} disk if the swing
amplifier was working. However, Meurer et al.\ (\markcite{mfbka99}1999) did not
detect any H$\alpha$ emission past $R_{\mbox{\scriptsize Ho}}$. Their
observations revealed only three faint \ion{H}{2} regions near the Holmberg
radius. These can be ionized by single late O or
early A type stars and are much fainter than those typically used to trace star
formation in galaxies (as done by, e.g., Kennicutt \markcite{k89}1989).
Therefore, at this point, it is unlikely that dark matter distributed in a
disk can account for all the properties of NGC~2915 simultaneously.

Nevertheless, it would be interesting to see this type of argument, based on
Toomre's (\markcite{t64}1964) $Q$ parameter and the presence of a spiral
pattern, be used more often in spiral galaxies to test for the presence of
substantial amount of unseen disk matter (see, e.g., Quillen \& Pickering
\markcite{qp97}1997; see also Quillen \& Sarajedini \markcite{qs98}1998 for a
similar application to intermediate-redshift galaxies).
\subsection{Slowly Rotating Triaxial Dark Halos\label{sec:triaxdm}}
\nopagebreak
Simulations of hierarchical structure formation in cold dark matter (CDM)
universes predict that the dark halos around galaxies are triaxial in the
mean, with some tendency to prolate shapes (Frenk et al.\ \markcite{fwde}1988;
Warren et al.\ \markcite{wqsz92}1992; Dubinski \& Carlberg
\markcite{dc91}1991). The triaxiality and oblateness are slightly stronger in
the outer parts, but are almost independent of the halo mass and of the ratio
$V_r/\sigma_v$ (rotational velocity to velocity dispersion), indicating that
the halos are supported by anisotropic velocity dispersions. The total angular
momentum vector is nevertheless preferentially aligned with the minor axis of
the mass distribution at all radii. When dissipation is included, the halos
become more oblate while the flattening is unchanged (Dubinski
\markcite{d94}1994).

Observationally, the constraints on the shape of galactic halos remain weak.
Polar-ring galaxies (e.g.\ Sackett et al.\ \markcite{srjf94}1994) and the
flaring (e.g.\ Olling \markcite{o95}1995) and warping (e.g.\ Sparke
\markcite{sp84a}1984a) of the outer \ion{H}{1} disk of spirals are usually
used to constrain the three-dimensional structure of halos. Such studies often
indicate flattened and sometimes triaxial dark halos (see also Becquaert \&
Combes \markcite{bc97}1997; Olling \markcite{o96}1996) but are as yet
inconclusive.

We now consider the possibility that the dark halo of NGC~2915 is triaxial,
with its figure rotating slowly about the rotation axis of the \ion{H}{1}
disk, and that the \ion{H}{1} bar and spiral arms are due to the forcing of
the \ion{H}{1} disk by this massive, extended, and rotating triaxial halo. The
halo would then act much like a very massive, slowly rotating bar. This would
readily explain why there is a unique pattern speed for the bar and spiral
arms and why the bar ends so far from corotation (see \S~\ref{sec:natpat}).
The situation which we are proposing is similar to that of rotating weak oval
distortions in disks, which are able to maintain an open spiral pattern to
large radii (see, e.g., Hunter et al.\ \markcite{hbheg88}1988). This is in
contrast to a more conventional bar, where the quadrupole term decreases too
rapidly outside corotation (where the bar ends) to maintain a spiral pattern.
Hydrodynamical simulations of gaseous disks in large tumbling triaxial halos
would be very useful to test our suggestion and to study the evolution of
disks in such potentials.

The straight line of nodes of NGC~2915's warp (\markcite{mcbf96}MCBF96)
supports the suggestion of a triaxial halo. Displacements between the angular
momentum vectors of the inner and outer regions of disks arise naturally in
CDM models because of tidal torques. In an axisymmetric but flattened
potential, the line of nodes would wrap quickly, but in a triaxial potential,
the principal axes of the halo provide natural directions with which the warp
can align (see Binney \markcite{bi92}1992 for a review of warps). Furthermore,
an annulus of orbits circulating around the short axis of a triaxial halo
becomes vertically unstable if the figure rotates, providing a natural way to
excite a warp. However, this may happen too far out in the disk (outside
corotation for prograde orbits) to be of interest here (Binney
\markcite{bi78}1978, \markcite{bi81}1981; see also Sparke
\markcite{sp8ab}1984b).

While the studies mentioned above (Frenk et al.\ \markcite{fwde}1988; Warren
et al.\ \markcite{wqsz92}1992; Dubinski \& Carlberg \markcite{dc91}1991;
Dubinski \markcite{d94}1994) have shown that dark halos in CDM cosmological
simulations can be strongly triaxial and have well-defined angular momentum
properties (with flat rotation curves extending down to the core), only the
instantaneous shapes and rotational motions within the figures (streaming)
have been studied. As far as we are aware, there has been no systematic study
yet of the {\em figure} rotation of these triaxial halos.
\placefigure{fig:halorot}

Pfitzner (1999, in preparation) recently began a study of the figure rotation
of halos in a set of large dissipationless CDM structure formation
simulations. The halos shape distribution varies systematically with both
radius and mass, but for our current purposes it suffices to say that many
halos are significantly triaxial, as in other studies. Following the
orientation of each halo over time, Pfitzner finds that a subset of triaxial
halos exhibit steady figure rotation about their minor axis. The statistics of
how many halos in the simulations have figure rotation is confused by
numerical resolution issues, but the fraction is clearly significant. To
verify that these results are not due to transient phenomena, one of the
rotating halo candidates was extracted from the simulations at a time near the
final epoch ($\approx$13~Gyr), and it was evolved in isolation. The halo
displayed no systematic change of density or shape over a time of 5 Gyr.
Figure~\ref{fig:halorot} shows that the figure of this triaxial halo rotates
as a solid body with constant pattern speed over the entire time span. Hence,
we conclude that at least some halos in these CDM simulations do exhibit
genuine steady slow figure rotation over a Hubble time. However, because the
halos formed are much more massive than that of NGC~2915, these simulations
are not yet directly applicable to systems of that size. They do, however,
suggest that an underlying triaxial dark halo with figure rotation could make
an important contribution to the dynamics of the outer \ion{H}{1} disk of some
galaxies. If this is correct, it also offers another possibility to measure
the shape of (triaxial) dark halos from detailed \ion{H}{1} velocity field,
using methods similar to those of Franx, van Gorkom, \& de Zeeuw
\markcite{fvd94}(1994).

The slow pattern speeds found in the present study and in the above
simulations are interesting in the context of recent work by Miwa \& Noguchi
\markcite{mn98}(1998). They made $N$-body simulations of barred galaxies in
which the bars arose (1) spontaneously in marginally stable disks, and (2)
through the effect of a tidal interaction in more stable disks. The pattern
speeds of the spontaneous bars are fast, with corotation near the ends of the
bars, as generally expected in self-consistent systems. In the tidally induced
bars, however, the pattern speeds are mostly much slower, and the bars appear
to be limited by the ILR, rather than by corotation. In particular, any
relatively large tidal disturbance leads to a slow pattern speed, independent
of the original properties of the disk. These results may be relevant to the
pattern speeds of the triaxial dark halos formed in CDM simulations, in which
the triaxiality results at least partially from tidal effects during the
aggregation of the halos.

Although more speculative, it may also be that triaxial halos are an essential
ingredient in the formation of starbursts. We may then be witnessing in
NGC~2915 the gradual buildup of a bulge, as gas driven to the center by the
triaxial halo potential and the bar is slowly transformed into stars by the
central starburst. However, the red diffuse stellar population detected by
\markcite{mmc94}MMC94 argues for an even earlier burst of star formation.
Nevertheless, if all the \ion{H}{1} were turned into stars, NGC~2915 would
appear very much like a normal spiral galaxy, with an ``old'' bulge and a
young disk which obeys the Tully-Fisher relation (Tully \& Fisher
\markcite{tf77}1977; assuming $M/L_B=1\;M_{\mbox{\scriptsize
    \sun}}/L_{B,\mbox{\scriptsize \sun}}$ for the gas). In this picture, the
gaseous disk of NGC~2915 would be almost pristine.
\section{Conclusion\label{sec:conclusions}}
We used the method of Tremaine \& Weinberg \markcite{tw84}(1984) to measure
the disk pattern speed in the galaxy NGC~2915, using neutral hydrogen as the
tracer and radio synthesis data. NGC~2915 is a BCD galaxy with a very extended
\ion{H}{1} disk showing an open barred spiral morphology. It is also strongly
dominated by dark matter. Our measurements yield a pattern speed of
$8.0\pm2.4$~km~s$^{-1}$~kpc$^{-1}$ for $D=5.3$~Mpc. This pattern speed is
inconsistent with the general property of self-consistent barred disks that
corotation occurs just outside the end of the bar. However, it agrees well
with more recent models considering dark halos with high central densities.
Our adopted bar length puts corotation at more than 1.7 bar radii. We
considered the possibility that the bar and spiral arms have different pattern
speeds but demonstrated that this is unlikely.

Independent of the pattern speed measured, the existence of a bar and spiral
pattern in NGC~2915 is also hard to explain. NGC~2915 is isolated, so
gravitational interactions are unlikely to explain the strong pattern observed
in the \ion{H}{1} disk. Because the surface density in the disk is so low,
Toomre's \markcite{t81}(1981) swing amplification mechanism is also unable to
explain the origin of the structures. Furthermore, resonance ring theory
fails to predict successfully the position of the pseudo-rings seen in the
disk of NGC~2915, indicating that the spiral arms are not driven by the bar
itself.

This lead us to consider two scenarios, involving different dark matter
distributions, to explain both the structure and dynamics of NGC~2915's
\ion{H}{1} disk. Firstly, we considered dark matter in near-circular motion,
distributed in a disk, and following closely the distribution of \ion{H}{1},
as suggested by the similarity of the dark matter and \ion{H}{1} surface
density profiles. This does not appear to work: if the dark matter disk is
dense enough for the swing amplifier to function efficiently, then the disk
becomes unstable to axisymmetric disturbances and we would expect to see
evidence of active star formation, which is not observed. Second, we
considered the effects of dark matter distributed in a massive
pressure-supported halo with extended triaxiality, like the halos seen in CDM
cosmological simulations. In addition, we required the halo to have a slow
figure rotation. Such figure rotation is seen in the preliminary analysis of a
set of CDM simulations. These show triaxial halos with figure rotation
constant over many Gyr. This kind of model has the potential to explain the
structures seen in the disk of NGC~2915, through the torques exerted by the
slowly rotating triaxial figure of the dark halo.

The structure in the \ion{H}{1} disk of NGC~2915 is visible because the radio
synthesis observations were made at a relatively high spatial resolution (beam
FWHM of about 650~pc for the uniformly weighted data and about 1150~pc for the
naturally weighted data). As far as we are aware, NGC~2915 is the only galaxy
showing such structure in an extended \ion{H}{1} disk. If our suggestions are
correct, it would be very desirable to have \ion{H}{1} observations of similar
spatial resolution for other nearby galaxies with extended \ion{H}{1} disks,
as they may also provide similar insight into the dynamics of their dark
halos. Our argument on figure rotation also relies heavily on the theory of
forcing of structures by oval distortions, in a situation which is not
precisely analogous to that in NGC~2915. Hydrodynamical simulations of gas
disks in the potentials of triaxial halos with figure rotation would thus be
very welcome.
\acknowledgments
We thank A.\ J.\ Kalnajs for comments on the manuscript and D.\ Pfenniger, J.\ 
A.\ Sellwood, and V.\ P.\ Debattista for useful discussions. M.\ Bureau
acknowledges the support of an Australian DEETYA Overseas Postgraduate
Research Scholarship and a Canadian NSERC Postgraduate Scholarship during the
conduct of this research. The Digitized Sky Surveys were produced at the Space
Telescope Science Institute under U.S. Government grant NAG W-2166. The images
of these surveys are based on photographic data obtained using the Oschin
Schmidt Telescope on Palomar Mountain and the UK Schmidt Telescope. The plates
were processed into the present compressed digital form with the permission of
these institutions.
%

%
\clearpage
%
%
%
\figcaption[fig1.ps]{Meurer et al.\ (1996) naturally weighted total \ion{H}{1}
  intensity map, overlaid on an optical image of NGC~2915 (Digitized Sky
  Survey). The beam is $45\arcsec\times45\arcsec$. Contours levels are 2.5,
  7.5, 15, 25, 50, 75, and 90 percent of the peak \ion{H}{1} surface
  brightness of 3.1~Jy~beam$^{-1}$~km~s$^{-1}$
  ($N_{HI}=1.71\times10^{21}$~cm$^{-2}$).
\label{fig:h1+optical}}
%
%
\figcaption[fig2.ps]{Meurer et al.\ (1996) naturally weighted \ion{H}{1}
  velocity field, overlaid on an optical image of NGC~2915 (Digitized Sky
  Survey). The beam is $45\arcsec\times45\arcsec$, and the velocity resolution
  is 6.62~km~s$^{-1}$.  Contours are heliocentric velocities separated by
  10~km~s$^{-1}$. A few contours are identified on the map.
\label{fig:velocity+optical}}
%
%
\figcaption[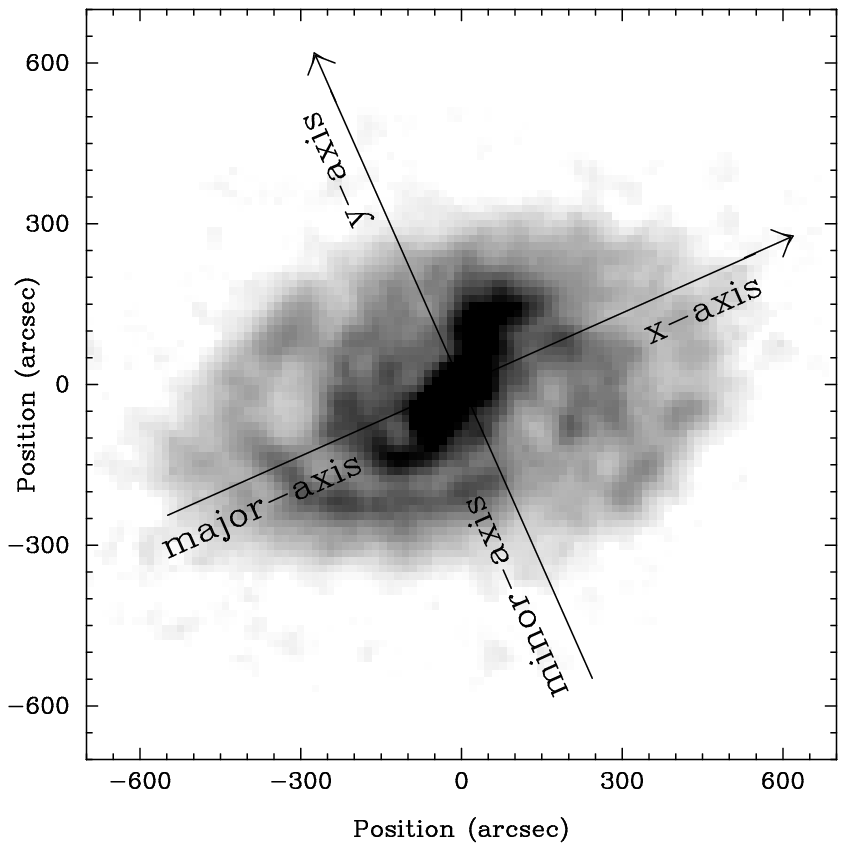]{Schematised view of the ``set-up'' needed for the
  Tremaine-Weinberg method calculations (eq.~[\ref{eq:tm}]), overlaid on a
  grayscale image of Meurer et al.\ (1996) naturally weighted total \ion{H}{1}
  intensity map ($45\arcsec\times45\arcsec$ beam).
\label{fig:h1+tm}} 
%
%
\figcaption[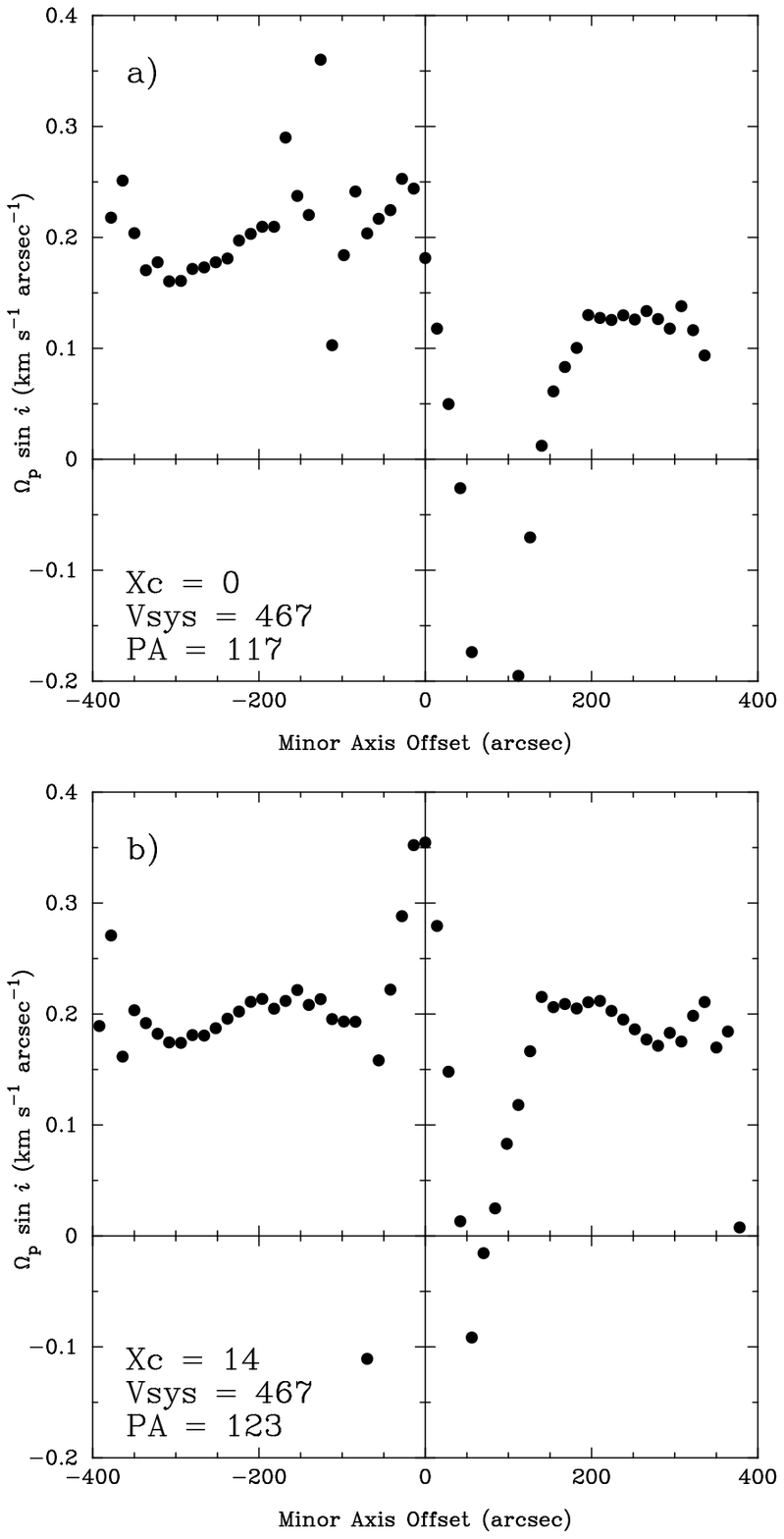]{Tremaine-Weinberg method measurements of the pattern
  speed in NGC~2915. (a) shows the case with Meurer et al.\ (1996) tilted-ring
  model best fit parameters, while (b) shows a typical good case. Parameters
  are indicated in the bottom left corner of each graph.
\label{fig:tmresults}}
%
%
\figcaption[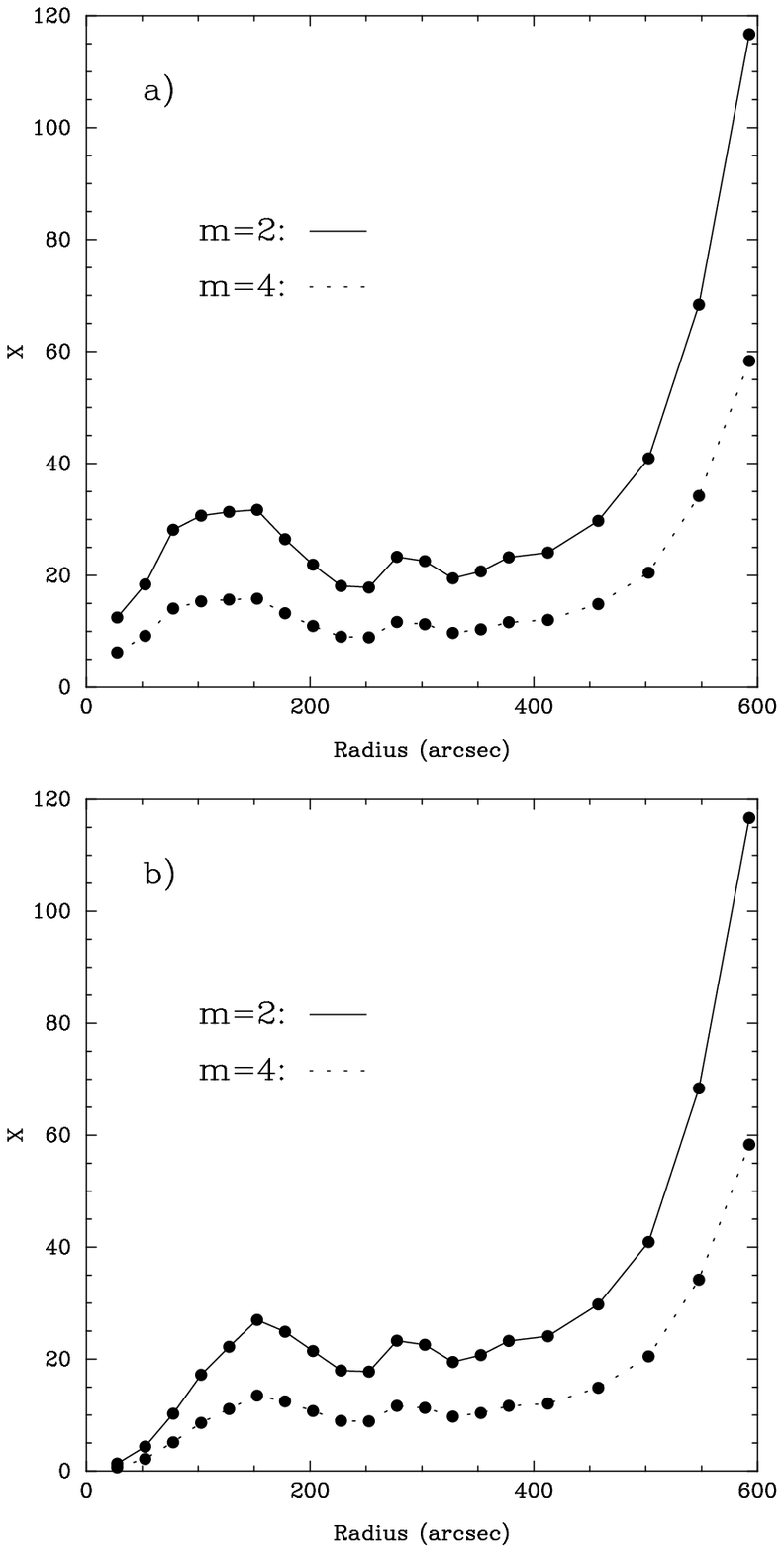]{Calculated swing amplifier parameter $X(r)$
  (eq.~[\ref{eq:Xr}]) for two-armed ($m=2$, full lines) and four-armed ($m=4$,
  dotted lines) spiral patterns in NGC~2915. (a) shows $X(r)$ when only the
  \ion{H}{1} distribution is taken into account, while in (b) both the
  \ion{H}{1} and the stellar component are considered. $X\lesssim3$ is
  required for swing amplification.
\label{fig:Xr}}
%
%
\figcaption[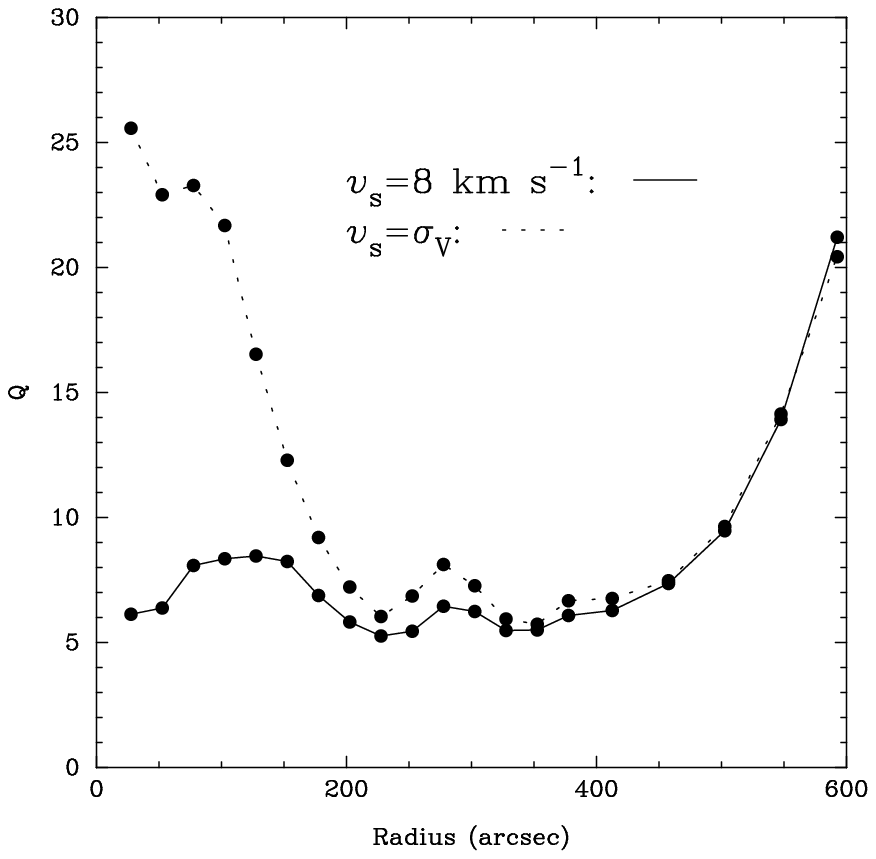]{Calculated stability parameter $Q(r)$ (eq.~[\ref{eq:Qr}])
  in NGC~2915. The full line shows $Q(r)$ considering a fixed sound speed
  $v_s(r)=8$~km~s$^{-1}$, equal to the velocity dispersion in the outer parts
  of the \ion{H}{1} disk, while the dotted line assumes $v_s(r)=\sigma_v(r)$.
  $Q(r)$ scales directly with $v_s(r)$. $Q\gtrsim2$ curbs swing amplification.
\label{fig:Qr}}
%
%
\figcaption[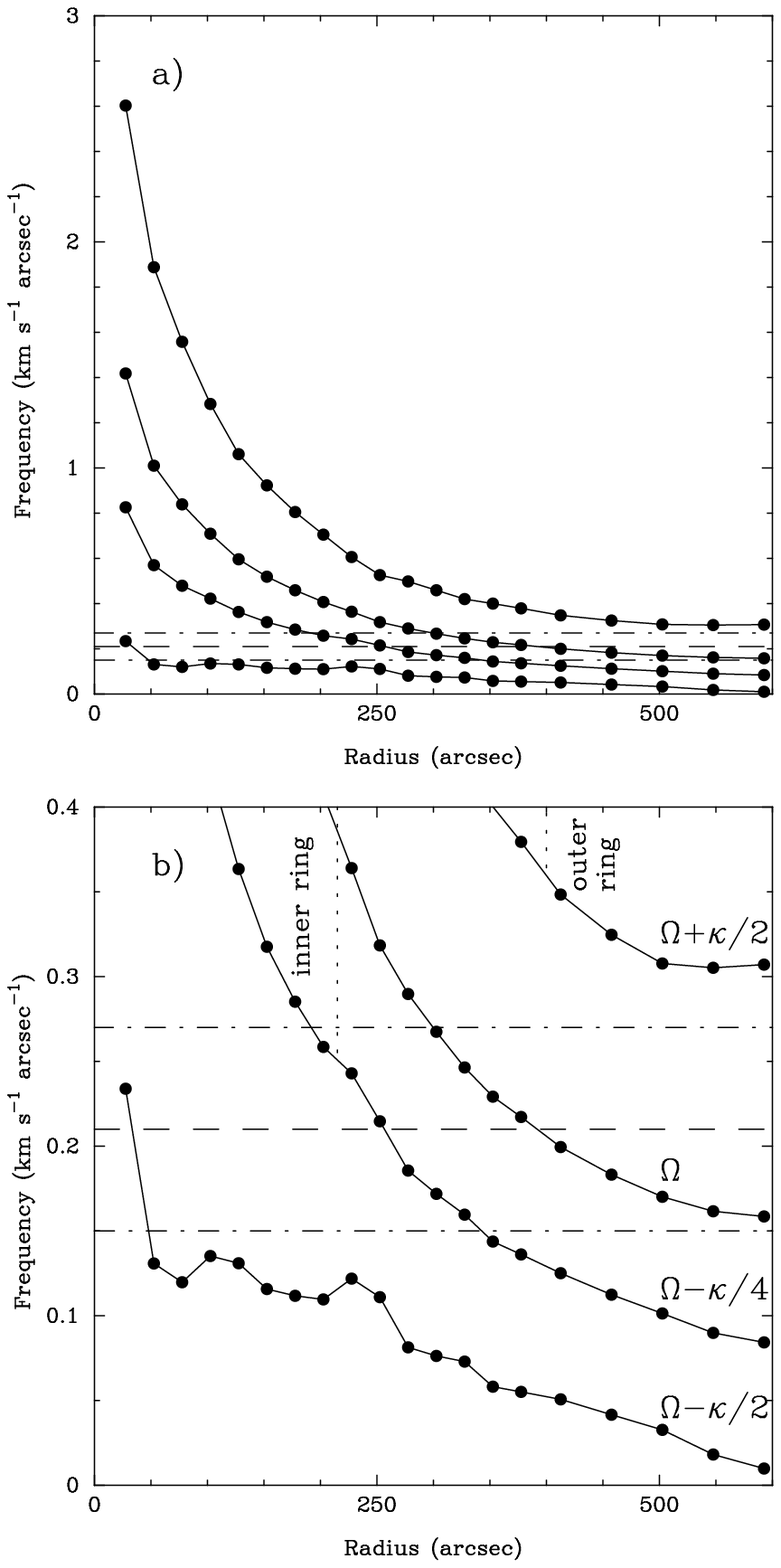]{(a) Calculated Lindblad precession frequencies for
  NGC~2915 (solid lines). From top to bottom, they are $\Omega+\kappa/2$,
  $\Omega$, $\Omega-\kappa/4$, and $\Omega-\kappa/2$ respectively. The dashed
  line shows the adopted pattern speed $\Omega_p$, while the dot-dashed lines
  show the range of values allowed by the uncertainty on $\Omega_p$. (b) is
  simply an enlargement of (a), but in addition the position of various
  features in the disk of NGC~2915 are indicated (dotted vertical lines).
\label{fig:resonances}}
%
%
\figcaption[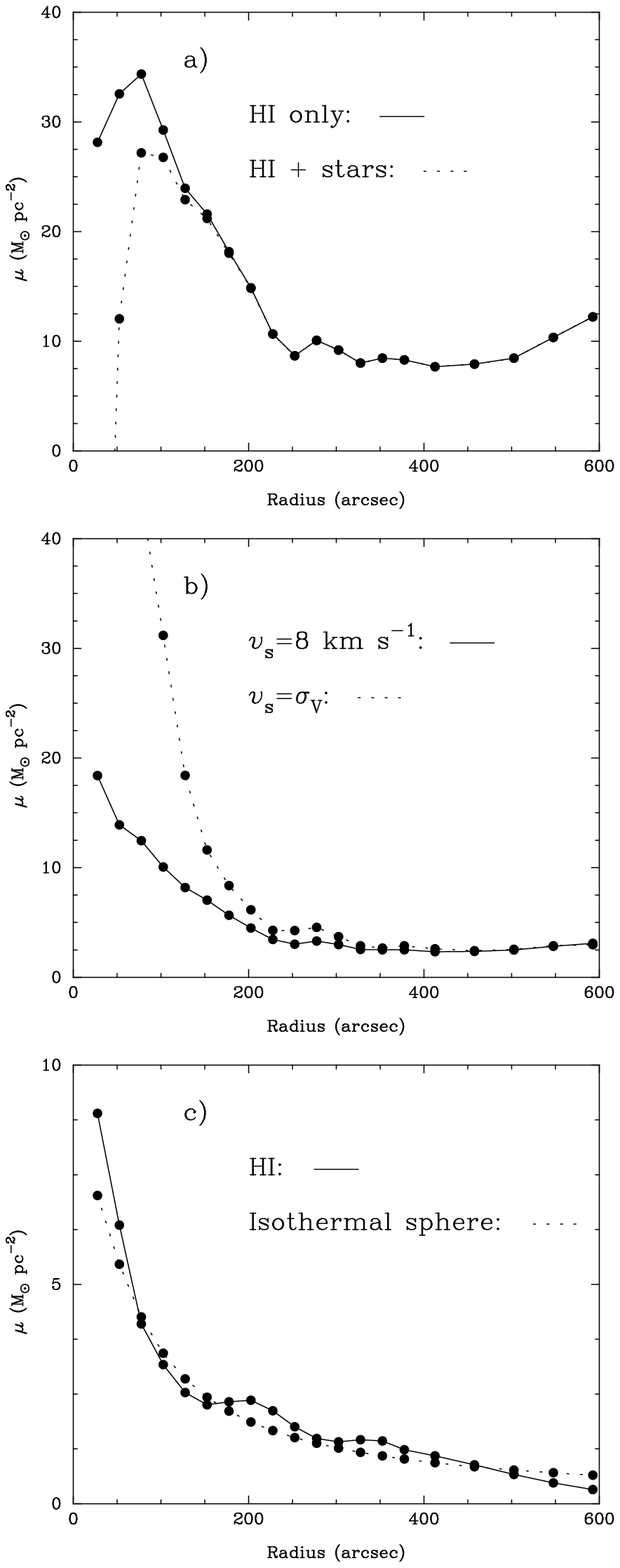]{(a) Calculated disk dark matter surface density needed to
  make the swing amplifier parameter $X(r)=3$ (eq.~[\ref{eq:Xr}]) for a
  two-armed ($m=2$) spiral pattern in NGC~2915. The full line shows $\mu_{DM}$
  when only the \ion{H}{1} distribution is taken into account, while for the
  dotted line both the \ion{H}{1} and the stellar component are considered.
  (b) Calculated disk dark matter surface density needed to make the stability
  parameter $Q(r)=2$ (eq.~[\ref{eq:Qr}]). The full line shows $\mu_{DM}$
  considering a fixed sound speed $v_s(r)=8$~km~s$^{-1}$, while the dotted
  line assumes $v_s(r)=\sigma_v(r)$. $Q(r)$ scales directly with $v_s(r)$. (c)
  The full line shows $\mu_{\mbox{\scriptsize HI}}$, the observed disk surface
  density of \ion{H}{1} (Table~\ref{ta:curves}), while the dotted line shows
  the projected dark matter surface density $\mu_{DM}$ of the dark isothermal
  sphere needed to model the rotation curve (model D of MCBF96, rescaled by 47.7).
\label{fig:dmprof}}
%
%
\figcaption[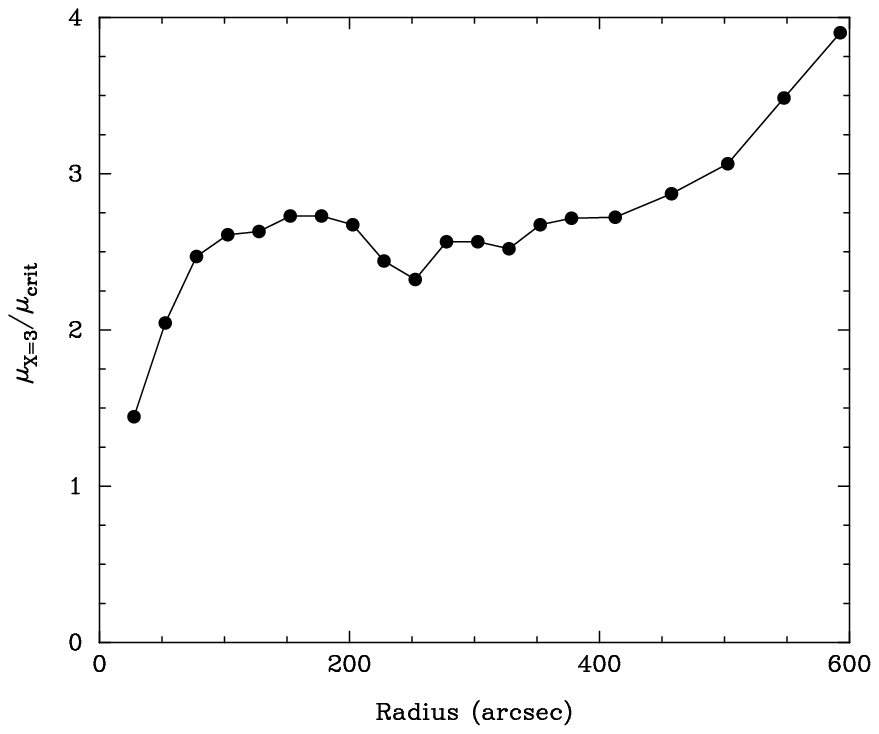]{Calculated ratio of the total surface density required to
  make the swing amplifier parameter $X(r)=3$ (eq.~[\ref{eq:Xr}] with $m=2$)
  to the critical surface density required for star formation in galactic
  disks (Kennicutt 1989).
\label{fig:muratio}}
%
%
\figcaption[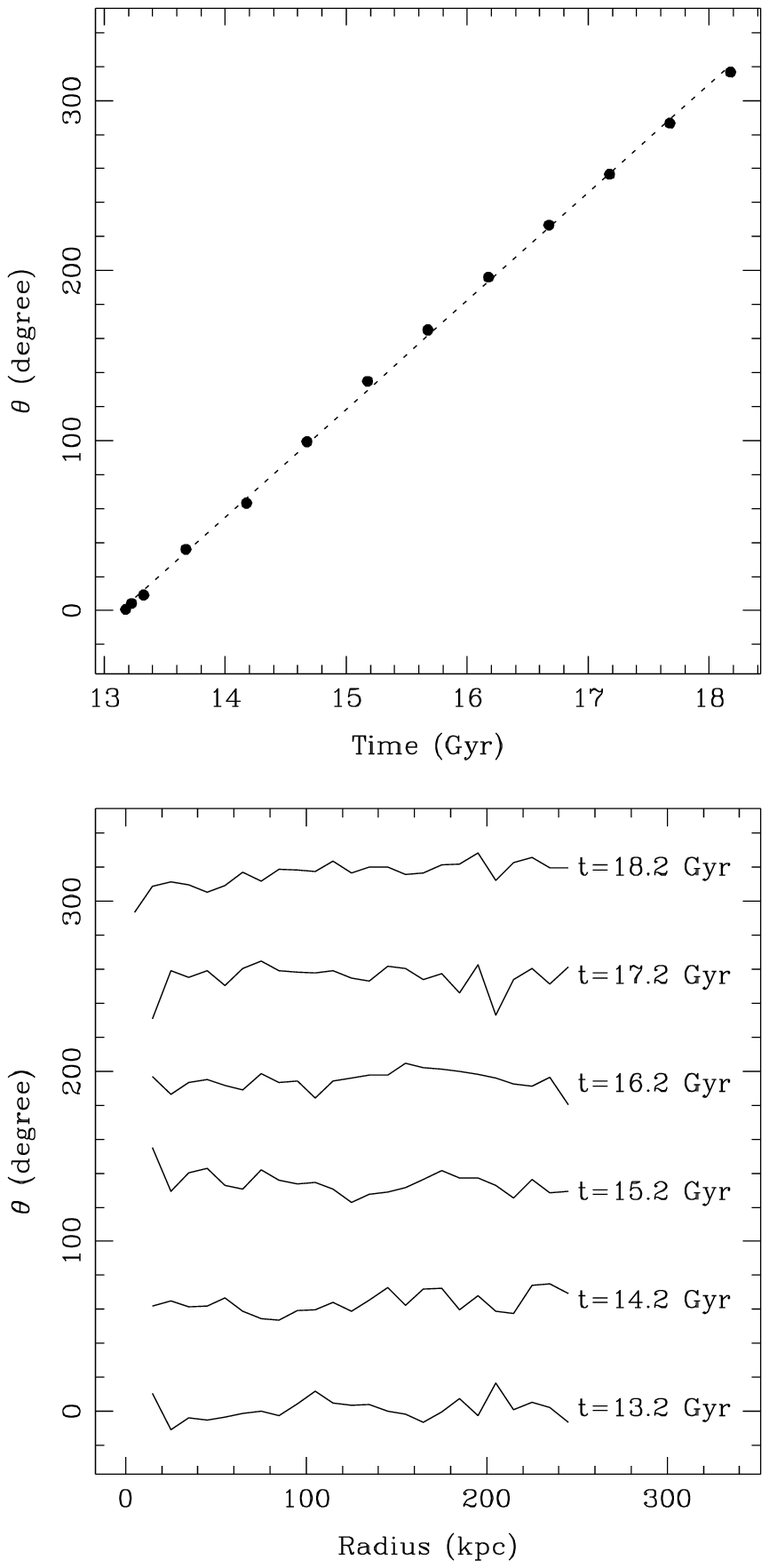]{(a) Orientation of the major axis of a dark halo formed
  in a CDM simulation and evolved in isolation (see \S~\ref{sec:triaxdm}) as a
  function of time. The angle $\theta$ is measured in a fixed plane
  perpendicular to the minor axis (the rotation axis). Filled circles show the
  mean orientation of the halo as a function of time, and the dotted line is a
  linear fit to the data. (b) Orientation of the major axis of the halo as a
  function of radius (in isodensity-defined shells) for selected times,
  showing that the halo rotates as a solid-body.
\label{fig:halorot}}
\clearpage
%
%
%
%
\begin{deluxetable}{llc}
\tablewidth{0pt}
\tablecaption{Basic Properties of NGC~2915 \label{ta:basic}}
\tablehead{\colhead{Quantity} & \colhead{Value} & \colhead{Reference}}
\startdata
Position (J2000) & $\alpha=9^{\mbox{\scriptsize h}} 26^{\mbox{\scriptsize m}} 11\fs83$ & 1 \nl
Position (J2000) & $\delta=-76\arcdeg 37\arcmin 35\farcs8$ & 1 \nl
Heliocentric Velocity & $V_{\mbox{\scriptsize \sun}}=468\pm5$~km~s$^{-1}$ & 2,3 \nl
Distance & $D=5.3\pm1.6$~Mpc & 1 \nl
Classification & I0/BCD & 1 \nl
Total Apparent $B$ Magnitude & $m_{B_T}=13.34$~mag & 1 \nl
Total Absolute $B$ Magnitude & $M_{B_T}=-15.90$~mag & 1 \nl
$B$ Disk Scalelength & $\alpha_B^{-1}=25\farcs6$ (660~pc) & 1 \nl
Corrected $B$ Central Surface Brightness & $B(0)_c=22.44$~mag~arcsec$^{-2}$ & 1 \nl
$B$ Holmberg Radius ($\mu_{B,0}=26.6$~mag~arcsec$^{-2}$) & $R_{\mbox{\scriptsize Ho},B}=1\farcm9$~(2.93~kpc) & 1 \nl
Total \ion{H}{1} Flux & $F_{\mbox{\scriptsize HI}}=145$~Jy~km~s$^{-1}$ & 2 \nl
Total \ion{H}{1} Mass & $M_{\mbox{\scriptsize HI}}=9.58\times10^8\;M_{\mbox{\scriptsize \sun}}$ & 2 \nl
\ion{H}{1} Velocity Width & $W_{20}=170$~km~s$^{-1}$ & 2 \nl
\ion{H}{1} Extent ($N_{\mbox{\scriptsize HI},0}=5\times10^{19}$~cm$^{-2}$) & $R_{\mbox{\scriptsize HI}}=9\farcm7$~(14.9~kpc) & 2 \nl
Dust Mass & $M_{\mbox{\scriptsize Dust}}=1.05\times10^4\;M_{\mbox{\scriptsize \sun}}$ & 3 \nl
Total Mass (within $r\approx600\arcsec$~=~15.4~kpc) & $M_{\mbox{\scriptsize Tot}}=2.7\times10^{10}\;M_{\mbox{\scriptsize \sun}}$ & 2 \nl
\enddata
\tablerefs{(1) Meurer et al.\ 1994; (2) Meurer et al.\ 1996; (3) Schmidt \&
Boller 1992.}
\end{deluxetable}
\clearpage
%
%
%
{\scriptsize
\begin{deluxetable}{ccccccc}
\tablewidth{0pt}
\tablecaption{Structural and Kinematical Profiles of NGC~2915 \label{ta:curves}}
\tablehead{\colhead{\coltwo{$r$}{(arcsec)}} & 
           \colhead{\coltwo{$\mu_{\mbox{\scriptsize HI}}$}{($M_{\mbox{\scriptsize \sun}}$~pc$^{-2}$)}} &
           \colhead{\coltwo{$\mu_B$}{($B$ mag arcsec$^{-2}$)}} &
           \colhead{\coltwo{$V_c$}{(km~s$^{-1}$)}} & 
           \colhead{\coltwo{$\sigma_v$}{(km~s$^{-1}$)}} & 
           \colhead{\coltwo{$\Omega$}{(km~s$^{-1}$~arcsec$^{-1}$)}} & 
           \colhead{\coltwo{$\kappa$}{(km~s$^{-1}$~arcsec$^{-1}$)}}}
\startdata
27.5  & 8.90 & 22.55& 39.0 & 33.4 & 1.42 & 2.37 \nl
52.5  & 6.35 & 23.97 & 53.0 & 28.8 & 1.01 & 1.76 \nl
77.5  & 4.10 & 25.11 & 65.0 & 23.0 & 0.84 & 1.44 \nl
102.5 & 3.17 & 26.26 & 72.7 & 20.8 & 0.71 & 1.15 \nl
127.5 & 2.53 & 27.20\tablenotemark{*} & 76.0 & 15.6 & 0.60 & 0.93 \nl
152.5 & 2.26 & 28.26\tablenotemark{*} & 79.2 & 11.9 & 0.52 & 0.81 \nl
177.5 & 2.32 & 29.32\tablenotemark{*} & 81.4 & 10.7 & 0.46 & 0.69 \nl
202.5 & 2.36 & 30.38\tablenotemark{*} & 82.5 & 9.9  & 0.41 & 0.60 \nl
227.5 & 2.12 & 31.44\tablenotemark{*} & 82.8 & 9.2  & 0.36 & 0.48 \nl
252.5 & 1.75 & 32.50\tablenotemark{*} & 80.4 & 10.1 & 0.32 & 0.41 \nl
277.5 & 1.49 & 33.56\tablenotemark{*} & 80.4 & 10.1 & 0.29 & 0.42 \nl
302.5 & 1.41 & 34.62\tablenotemark{*} & 80.9 & 9.3  & 0.27 & 0.38 \nl
327.5 & 1.46 & 35.68\tablenotemark{*} & 80.7 & 8.7  & 0.25 & 0.35 \nl
352.5 & 1.43 & 36.74\tablenotemark{*} & 80.8 & 8.3  & 0.23 & 0.34 \nl
377.5 & 1.23 & 37.80\tablenotemark{*} & 82.0 & 8.8  & 0.22 & 0.32 \nl
412.5 & 1.09 & 39.28\tablenotemark{*} & 82.3 & 8.6  & 0.20 & 0.30 \nl
457.5 & 0.89 & 41.19\tablenotemark{*} & 83.8 & 8.1  & 0.18 & 0.28 \nl
502.5 & 0.67 & 43.10\tablenotemark{*} & 85.5 & 8.1  & 0.17 & 0.28 \nl
547.5 & 0.48 & 45.01\tablenotemark{*} & 88.5 & 8.1  & 0.16 & 0.29 \nl
592.5 & 0.32 & 46.92\tablenotemark{*} & 93.9 & 7.7  & 0.16 & 0.30 \nl
\enddata
\tablenotetext{*}{Extrapolated values from the exponential fit to the outer
  parts of the disk of MMC94.}
\end{deluxetable}
}
%
%
\clearpage
%
%
%
%
%
%
%
%
\begin{figure}
\epsscale{0.6}
\plotone{fig3.ps}
\end{figure}
\clearpage
%
%
\begin{figure}
\epsscale{0.6}
\plotone{fig4.ps}
\end{figure}
\clearpage
%
%
\begin{figure}
\epsscale{0.6}
\plotone{fig6.ps}
\end{figure}
\clearpage
%
%
\begin{figure}
\epsscale{0.6}
\plotone{fig7.ps}
\end{figure}
\clearpage
%
%
\begin{figure}
\epsscale{0.6}
\plotone{fig5.ps}
\end{figure}
\clearpage
%
%
\begin{figure}
\epsscale{0.5}
\plotone{fig8.ps}
\end{figure}
\clearpage
%
%
%
\begin{figure}
\epsscale{0.6}
\plotone{fig9.ps}
\end{figure}
\clearpage
%
%
%
\begin{figure}
\epsscale{1.0}
\plotone{fig10.ps}
\end{figure}
\clearpage

\begin{references}
%
\reference{ani79} Aoki, S., Noguchi, M, \& Iye, M.\ 1979, \pasj, 31, 737
\reference{a92a} Athanassoula, E.\ 1992a, \mnras, 259, 328
\reference{a92b} ---------.\ 1992b, \mnras, 259, 345
\reference{abp87} Athanassoula, E., Bosma, A., \& Papaioannou, S.\ 1987, \aap,
179, 23
\reference{as86} Athanassoula, E., \& Sellwood, J.\ A.\ 1986, \mnras, 221, 213
\reference{b75} Bardeen, J.\ M.\ 1975, in Dynamics of Stellar Systems, ed.\
A.\ Hayli (Reidel), 297
\reference{bc97} Becquaert, J.-F., \& Combes, F.\ 1997, \aap, 325, 41
\reference{b89} Begeman, K.\ G.\ 1989, \aap, 223, 47
\reference{bi78} Binney, J.\ J.\ 1978, \mnras, 183, 779
\reference{bi81} ---------.\ 1981, \mnras, 196, 455
\reference{bi92} ---------.\ 1992, \araa, 30, 51
\reference{b78} Bosma, A.\ 1978, Ph.D.\ Thesis, University of Groningen
\reference{b81} ---------.\ 1981, \aj, 86, 1971
\reference{b92} Broeils, A.\ H.\ 1992, \aap, 256, 19
\reference{bod98} Byrd, G.\ G., Ousley, D., \& Dalla Piazza C.\ 1998, \mnras, 298, 78
\reference{brsbc94} Byrd, G., Rautiainen, P., Salo, H., Buta, R., \& Crocker,
D.\ A.\ 1994, \aj, 108, 476
\reference{ca93} Canzian, B.\ 1993, \apj, 414, 487
\reference{cb89} Carignan, C., \& Beaulieu, S.\ 1989, \apj, 347, 760
\reference{cf85} Carignan, C., \& Freeman, K.\ C.\ 1985, \apj, 294, 494
\reference{c94} Carr, B.\ 1994, \araa, 32, 531
\reference{c93} Combes, F.\ 1993, in $N$-Body Problems and Gravitational Dynamics,
ed.\ F.\ Combes \& E.\ Athanassoula (Paris: Observatoire de Paris), 137
\reference{cp97} Combes, F., \& Pfenniger, D.\ 1997, \aap, 327, 453
\reference{cs81} Combes, F., \& Sanders, R.\ H.\ 1981, \aap, 96, 164
\reference{c80} Contopoulos, G.\ 1980, \aap, 81, 198
\reference{cg89} Contopoulos, G., \& Grosb\o l, P.\ 1989, \aapr, 1, 261
\reference{ds98} Debattista, V.\ P., \& Sellwood, J.\ A.\ 1998, \apj, 493, L5
\reference{d94} Dubinski, J.\ 1994, \apj, 431, 617
\reference{dc91} Dubinski, J., \& Carlberg, R.\ G.\ 1991, \apj, 378, 496
\reference{e96} Elmegreen, B.\ 1996, in Barred Galaxies, ed.\ R.\ Buta, D.\
A.\ Crocker, \& B.\ G.\ Elmegreen (San Francisco:ASP), 197
\reference{fvd94} Franx, M., van Gorkom, J.\ H., \& de Zeeuw, T.\ 1994, \apj,
436, 642
\reference{fwde} Frenk, C.\ S.\, White, S.\ D.\ M., Davis, M., \& Efstathiou,
G.\ 1988, \apj, 327, 507
\reference{gca90} Gerin, M., Combes, F, \& Athanassoula, E.\ 1990, \aap, 230,
37
\reference{gkm99} Gerssen, J., Kuijken, K., \& Merrifield, M.\ R.\ 1999,
\mnras, submitted.
\reference{gl65} Goldreich, P., \& Lynden-Bell, D.\ 1965, \mnras, 130, 125
\reference{hbheg88} Hunter, J.\ H., Ball, R., Huntley, J.\ M., England, M.\
N., \& Gottesman, S.\ T.\ 1988, \apj, 324, 721 
\reference{k71} Kalnajs, A.\ J.\ 1971, \apj, 166, 275
\reference{k77} ---------.\ 1977, \apj, 212, 637
\reference{k89} Kennicutt, R.\ C.\ Jr.\ 1989, \apj, 344, 685 
\reference{k87} Kent, S.\ M.\ 1987, \aj, 93, 1062
\reference{k90} ---------.\ 1990, \aj, 100, 377
\reference{k84} Kormendy, J.\ 1984, \apj, 286, 132
\reference{km96} Kuijken, K., \& Merrifield, M.\ R.\ 1996, in Barred Galaxies,
ed.\ R.\ Buta, D.\ A.\ Crocker, \& B.\ G.\ Elmegreen (San Francisco:ASP), 215
\reference{la93} Lequeux, J., \& Allen, R.\ J.\ 1993, \aap, 280, L23 
\reference{lk96} Lindblad, P.\ A.\ B., \& Kristen, H.\ 1996, \aap, 313, 733
\reference{lla96} Lindblad, P.\ A.\ B., Lindblad, P.\ O., \& Athanassoula,
E. 1996, \aap, 313, 65 
\reference{mhws95} Marlowe, A.\ T., Heckman, T.\ M., Wyse, R.\ F.\ G., \&
Schommer, R.\ 1995, \apj, 438, 563 
\reference{mk95} Merrifield, M.\ R., \& Kuijken K.\ 1995, \mnras, 274, 933
\reference{mcbf96} Meurer, G.\ R., Carignan, C., Beaulieu, S.\ F., \& Freeman,
K.\ C.\ 1996, \aj, 111, 1551 (MCBF96)
\reference{mfbka99} Meurer, G.\ R., Freeman, K.\ C., Bland-Hawthorn, J.,
Knezek, P.\ M., \& Allen, R.\ J.\ 1999, \baas, 31, 828
\reference{mmc94} Meurer, G.\ R., Mackie, G., \& Carignan, C.\ 1994, \aj, 107,
2021 (MMC94)
\reference{mn98} Miwa, T., \& Noguchi, M.\ 1998, \apj, 499, 149
\reference{n87} Noguchi, M.\ 1987, \mnras, 228, 635
\reference{o95} Olling, R.\ P.\ 1995, \aj, 110, 591
\reference{o96} ---------.\ 1996, \aj, 112, 481
\reference{pp86} Petrou, M., \& Papayannopoulos, T.\ 1986, \mnras, 219, 157
\reference{pc94} Pfenniger, D., \& Combes, F. 1994, \aap, 285, 94
\reference{pcm94} Pfenniger, D., Combes, F., \& Martinet, L. 1994, \aap, 285, 79 
\reference{qs98} Quillen, A.\ C., \& Sarajedini, V.\ L.\ 1998, \aj, 115, 1412
\reference{qp97} Quillen, A.\ C., \& Pickering, T.\ E.\ 1997, \aj, 113, 2075
\reference{rsjk91} Raha, N., Sellwood, J.\ A., James, R.\ A., \& Kahn, F.\ D.\
1991, \nat, 352, 411
\reference{rbtssw96} Ryder, S.\ D., Buta, R.\ J., Toledo, H., Shukla, H.,
Staveley-Smith, L., \& Walsh, W.\ 1996, \apj, 460, 665
\reference{srjf94} Sackett, P.\ D., Rix, H.-W., Jarvis, B.\ J., \& Freeman,
K.\ C.\ 1994, \apj, 436, 629
\reference{sas79} Sancisi, R., Allen, R.\ J., \& Sullivan, W.\ T.\ 1979, \aap,
78, 217
\reference{sb92} Schmidt, K.-H., \& Boller, T.\ 1992, ANac, 313, 189
\reference{s81} Schwarz, M.\ P.\ 1981, \apj, 247, 77
\reference{s84} ---------.\ 1984, \mnras, 209, 93
\reference{sj98} Seiger, M.\ S., \& James, P.\ A.\ 1998, \mnras, 299, 672
\reference{s80} Sellwood, J.\ A.\ 1980, \aap, 89, 296
\reference{se81} ---------\ 1981, \aap, 99, 362
\reference{ss88} Sellwood, J.\ A., \& Sparke, L.\ S.\ 1988, \mnras, 231, 25
\reference{sw93} Sellwood, J.\ A., \& Wilkinson, A.\ 1993, Re.\ Prog.\ Phys.,
56, 173 
\reference{sp84a} Sparke, L.\ S.\ 1984a, \apj, 280, 117
\reference{sp84b} ---------.\ 1984b, \mnras, 211, 911
\reference{ss87} Sparke, L.\ S., \& Sellwood, J.\ A.\ 1987, \mnras, 225, 653
\reference{stap88} Sygnet, J.\ F., Tagger, M., Athanassoula, A., \& Pellat, R.\
1988, \mnras, 232, 733
\reference{tsap87} Tagger, M., Sygnet, J.\ F., Athanassoula, A., \& Pellat, R.\
1987, \apj, 318, L43
\reference{ts85} Teuben, P.\ J., \& Sanders, R.\ H.\ 1985, \mnras, 212, 257 
\reference{t64} Toomre, A.\ 1964, \apj, 139, 1217
\reference{t81} ---------.\ 1981, in Structure and Evolution of Normal Galaxies,
ed.\ S.\ M.\ Fall \& D.\ Lynden-Bell (Cambridge: Cambridge University Press),
111 
\reference{tm84} Tremaine, S., \& Weinberg, M.\ D.\ 1984, \apjl, 282, L5
\reference{t87} Trimble, V.\ 1987, \araa, 25, 425
\reference{tf77} Tully, R.\ B., \& Fisher, J.\ R.\ 1977, \aap, 54, 661 
\reference{wqsz92} Warren, M.\ S., Quinn, P.\ J., Salmon, J.\ K., \& Zurek,
W.\ H.\ 1992, \apj, 399, 405
\reference{w85} Weinberg, M.\ D.\ 1985, \mnras, 213, 451
\reference{wm94} Wilson, T.\ L., \& Mauersberger, R.\ 1994, \aap, 282, L41
%
\end{references}
\end{document}